\documentclass[a4paper,fleqn,usenatbib]{mnras}

\usepackage{mathptmx}

\usepackage{ae,aecompl}

\usepackage{graphicx}	
\usepackage{amsmath}	
\usepackage{amssymb}	

\usepackage{color,epsfig} 
\usepackage[utf8]{inputenc} 

\usepackage{multirow}

\usepackage[flushleft]{threeparttable}
\makeatletter
\newlength{\abovecaptionskip}%
\makeatother

\newcommand{\be}{\begin{equation}}
\newcommand{\ee}{\end{equation}}

\newcommand*\diff{\mathop{}\!\mathrm{d}}
\newcommand{\vect}[1]{\mathbf{#1}}

\newcommand{\mstar}{M_{\star}}
\newcommand{\rstar}{R_{\star}}

\newcommand{\lstar}{\ell_{\star}}

\newcommand{\rp}{R_{\rm p}}
\newcommand{\rt}{R_{\rm t}}

\newcommand{\rint}{R_{\rm int}}
\newcommand{\racc}{R_{\rm acc}}

\newcommand{\rb}{R_{\rm b}}

\newcommand{\mdot}{\dot{M}}
\newcommand{\vn}{v_{\rm n}}
\newcommand{\ve}{v_{\rm e}}

\newcommand{\mpart}{M_{\rm p}}

\newcommand{\mh}{M_{\rm h}}

\newcommand{\tmin}{t_{\rm min}}
\newcommand{\amin}{a_{\rm min}}

\newcommand{\ex}{\vect{e}_{\rm x}}
\newcommand{\ey}{\vect{e}_{\rm y}}
\newcommand{\ez}{\vect{e}_{\rm z}}

\newcommand{\er}{\vect{e}_{\rm r}}

\newcommand{\rg}{R_{\rm g}}

\newcommand{\mdotout}{\dot{M}_{\rm out}}
\newcommand{\mdotin}{\dot{M}_{\rm in}}

\newcommand{\mdotacc}{\dot{M}_{\rm acc}}
\newcommand{\mdotp}{\dot{M}_{\rm p}}
\newcommand{\mdotedd}{\dot{M}_{\rm Edd}}
\newcommand{\mdotfb}{\dot{M}_{\rm fb}}

\def\m6{\, M_6}

\def\a3{\, a_3}

\def\msun{\, \mathrm{M}_{\hbox{$\odot$}}}
\def\rsun{\, \mathrm{R}_{\hbox{$\odot$}}}
\def\gcm3{\, \rm g \, cm^{-3}}

\def\days{\, \rm d}

\def\ergpers{\, \rm erg\, s^{-1}}
\def\msunyr{\, \rm \mathrm{M}_{\hbox{$\odot$}} \, {yr}^{-1}}
\def\cm2g{\, \rm cm^{2} \, g^{-1}}

\title[Disc formation in TDEs]{Simulating realistic disc formation in tidal disruption events}

\author[C. Bonnerot and W. Lu]{Clément Bonnerot$^{1}$\thanks{E-mail: bonnerot@tapir.caltech.edu} and Wenbin Lu$^{1}$\thanks{E-mail: wenbinlu@caltech.edu}
\\
$^{1}$TAPIR, Mailcode 350-17, California Institute of Technology, Pasadena, CA 91125, USA\\
}
\date{Accepted XXX. Received YYY; in original form ZZZ}

\pubyear{2019}

\begin{document}
\label{firstpage}
\pagerange{\pageref{firstpage}--\pageref{lastpage}}
\maketitle

\begin{abstract}

A star coming too close to a supermassive black hole gets disrupted by the tidal force of the compact object in a tidal disruption event, or TDE. Following this encounter, the debris evolves into an elongated stream, half of which coming back to pericenter. Relativistic apsidal precession then leads to a self-crossing shock that initiates the formation of an accretion disc. We perform the first simulation of this process considering a realistic stellar trajectory and black hole mass, which has so far eluded investigations for computational reasons. This numerical issue is alleviated by using as initial conditions the outflow launched by the self-crossing shock according the local simulation of \citet{lu2019}. We find that the gas leaving the intersection point experiences numerous secondary shocks that result in the rapid formation of a thick and marginally-bound disc. The mass distribution features two overdensities identified as spiral shocks that drive slow gas inflow along the mid-plane. Inward motion primarily takes place along the funnels of the newly-formed torus, from which a fraction of the matter can get accreted. Further out, the gas moves outward forming an extended envelope completely surrounding the accretion flow. Secondary shocks heat the debris at a rate of a few times $10^{44} \ergpers$ with a large fraction likely participating to the bolometric luminosity. These results pave the way towards a complete understanding of the early radiation from TDEs that progressively becomes accessible from observations.

\end{abstract}





\section{Introduction}
\label{introduction}

Tidal disruption events (TDEs) occur when a star is disrupted by a supermassive black hole, after which the debris fuels the compact object producing a luminous multi-wavelength flare lasting from months to years \citep{rees1988}. These events represent unique tools to study black holes in the center of otherwise quiescent galaxies as well as their dust and gas environments. Furthermore, they hold promise for improving our understanding of multiple processes occurring around compact objects, such as super-Eddington accretion and the formation of relativistic jets. Because the victim star inherits its final trajectory from two-body interactions taking place in the surrounding nuclear stellar cluster, TDEs can also be used to study the various dynamical processes at play in this region.

The sample of TDE candidates is steadily growing as more of them are detected at  optical, UV and soft X-ray wavelengths \citep{bade1996,komossa1999-rxj1242,van_velzen2011,gezari2006,holoien2016-14li,gezari2017-15oi}. Thanks to the advent of high-cadence surveys, a larger fraction of these observed events have well-sampled lightcurves that include the early rise to peak \citep{gezari2012,chornock2014,arcavi2014,van_velzen2019,blagorodnova2019,holoien2019} emerging when the stellar debris start radiating at an observable level. A few of them also displays hard X-ray emission that is attributed to relativistic jets launched from the inner region of the accretion flow \citep{bloom2011,cenko2012-j2058,brown2015}.

The star is first scattered onto a plunging trajectory very close to parabolic after a series of stellar encounters taking place at large distances \citep{frank1976}. Its disruption by the black hole imparts a spread in orbital energy to the debris given by its depth inside the gravitational potential well \citep{evans1989,lodato2009,guillochon2013}. The gas subsequently follows a wide range of ballistic and highly eccentric trajectories to evolve into an elongated stream that remains thin owing to the confining effect of its self-gravity \citep{kochanek1994,guillochon2014-10jh,coughlin2016-structure}. Roughly half this debris is bound and thus falls back towards the black hole at a rate entirely specified by its orbital energy distribution \citep{rees1988,phinney1989}. A wide consensus has been reached regarding the hydrodynamics at play during these early stages of evolution and it is not strongly affected by the inclusion of additional physical ingredients \citep{coughlin2015-variability,kesden2012-lightcurve,tejeda2017,gafton2019,bonnerot2017-magnetic,steinberg2019}.

When the first debris reaches pericenter, it experiences relativistic apsidal precession that causes its collision with the part of the stream still approaching the compact object. This self-crossing shock has been investigated through local simulations \citep{kim1999,jiang2016,lu2019}, which find that it is close to adiabatic and therefore causes the formation of an outflow from the intersection point. This initial source of dissipation has been proposed to initiate the formation of an accretion disc, during which the gas trajectories progressively become more circular. This process has been investigated semi-analytically \citep{kochanek1994,dai2015,bonnerot2017-stream} but simulations to date have only been able to follow it for encounters with an intermediate-mass black hole \citep{rosswog2009,guillochon2014-10jh,shiokawa2015} or involving a star with an initially bound trajectory \citep{hayasaki2013,bonnerot2016-circ,hayasaki2016-spin,sadowski2016}. These conditions are adopted for entirely numerical reasons because they artificially decrease the length of the stream that alleviates the huge computational cost required to accurately resolve it in the realistic case of a parabolic encounter with a supermassive black hole. These numerical investigations led nevertheless to significant improvements in our understanding of how disc formation occurs. Most importantly, they put forward major issues with the early picture of a fast circularization into a compact ring located near the tidal radius \citep{rees1988}. It was found that the accretion flow extends to much larger distances and could require a few to many orbital timescales to assemble \citep{shiokawa2015,bonnerot2016-circ}. However, it remains uncertain how to quantitatively extrapolate the results of these simplified studies to the physically-motivated situation.

A clear understanding of the disc formation at play in real TDEs is paramount since this process connects the well-understood phases of stellar disruption and early stream evolution to the properties of the nascent accretion flow, from which most of the radiation is expected to emerge. For instance, it can inform the various models for the origin of the optical emission from these events that relies on either self-crossing shocks happening at large radii \citep{piran2015-circ} or reprocessing of radiation generated close to the tidal radius \citep{metzger2016}. The resulting gas distribution also affects the diffusion of photons produced near the black hole and whose interactions with the surrounding matter determines the spectrum of these events, in particular the expected strengths of the various lines \citep{roth2016}. Furthermore, the outcome of the disc formation process specifies the correct initial conditions to be used in simulations of gas accretion in this newly-formed structure where matter is thought to spiral inwards under the effect of magnetic stresses that can also be accompanied with the formation of relativistic jets \citep{dai2018,curd2018}.

In this paper, we carry out the first realistic simulation of disc formation following the disruption of a star on a parabolic orbit by a supermassive black hole. This is achieved using the smoothed-particle-hydrodynamics (SPH) method by injecting gas inside the computational domain according to the properties obtained from the local study of \citet{lu2019} for the outflow launched from the self-crossing shock. With this procedure, we are able to circumvent the high computational cost required to follow the entire evolution of the stream that has so far prohibited simulations of realistic disc formation in TDEs. Our simulation captures the complex interactions experienced by the shocked gas after leaving the intersection point that allowed us to follow the formation of the accretion flow as well as several other hydrodynamical phenomena.

This paper is organized as follows. In Section \ref{sec:simulations}, we explain the initial conditions and numerical setup used for our simulation as well as the injection procedure used to model the outflow launched from the self-crossing shock. The results are presented in Section \ref{sec:results} focusing on the disc assembly through secondary shocks, gas inflow and outflow as well as heating and radiation from the system. Section \ref{sec:discussion} includes a resolution study and a discussion of several physical effects that are not included in our simulation. Finally, we summarize our main findings in Section \ref{sec:summary}.

\section{SPH simulations}
\label{sec:simulations}

We simulate the evolution of the stellar debris after it has passed through the self-crossing shock. This is achieved numerically by continuously injecting gas from an outflowing box located at the intersection point according to the properties of the shocked gas obtained from the local simulation of \citet{lu2019}.

\subsection{Initial conditions}
\label{sec:initial-conditions}

As a representative example, we consider the disruption of a star with mass $\mstar = 0.5 \msun$ and radius $\rstar = 0.46 \rsun$ on a parabolic orbit by a black hole of mass $\mh = 2.5 \times 10^6 \msun$. The pericenter distance $\rp$ is equal to the tidal radius
\be
\rt = \rstar \left(\frac{\mh}{\mstar}\right)^{1/3} \approx 15 \, \rg = 5.5 \times 10^{12} \, \rm cm,
\label{eq:tidal-radius}
\ee
where $\rg = G \mh/c^2$ is the gravitational radius of the black hole. The corresponding penetration factor, defined as the ratio of tidal radius to pericenter distance, is $\beta =1$. Contrary to all previous numerical study of disc formation, this choice of parameters is physically motivated for a realistic disruption of a main sequence star by a supermassive black hole. Because we consider this already a significant step forward in the understanding of this process, we defer a more exhaustive exploration of the parameter space to future studies.

For this set of parameters, the stellar debris gets imparted an energy spread \citep{stone2013}
\be
\Delta \epsilon = \frac{G \mh \rstar}{\rt^2},
\ee
by the disruption, putting the most bound part on a orbit with semi-major axis
\be
\amin = \frac{G \mh}{2 \Delta \epsilon} = \frac{\rstar}{2} \left(\frac{\mh}{\mstar} \right)^{1/3} \approx 85 \rt.
\label{eq:semi-major-axis}
\ee
According to Kepler's third law, it corresponds to an orbital period
\be
\tmin = 2^{-1/2} \pi \left(\frac{G \mstar}{\rstar^3}\right)^{-1/2} \left(\frac{\mh}{\mstar}\right)^{1/2} \approx 40 \days,
\ee
which we will use as our reference time-scale throughout the paper. Note that the all gas elements keep instead an angular momentum close to that of the original star given by $\ell_{\star} = (2 G \mh \rt)^{1/2}$ for a parabolic orbit since the redistribution of this orbital parameter is negligible during the disruption process. As a result, the gas distribution evolves into an elongated stream, whose bound half comes back to the original stellar pericenter. This fallback of matter proceeds at a rate given by \citep{rees1988,phinney1989}
\be
\dot{M}_{\rm fb} = \dot{M}_{\rm p} \left(1+\frac{t}{\tmin}\right)^{-5/3},
\label{eq:fallback_rate}
\ee
where $t=0$ corresponds to the peak of the fallback rate given by 
\be
\dot{M}_{\rm p} = \frac{\mstar}{3 \tmin} \approx 25 \, \dot{M}_{\rm Edd} = 1.5 \msunyr ,
\ee
as obtained by setting to half a stellar mass the integral of equation \eqref{eq:fallback_rate} from 0 to infinity. The Eddington accretion rate is obtained from $\dot{M}_{\rm Edd} = L_{\rm Edd}/(\eta c^2) \approx 0.06 \msunyr$ for a radiative efficiency of $\eta = 0.1$ where $L_{\rm Edd} = 3.2 \times 10^{44} \ergpers$ denotes the Eddington luminosity. We note that equation \eqref{eq:fallback_rate} assumes a flat energy distribution within the debris for simplicity keeping in mind that the associated decay law for the fallback rate varies when the density structure of the star is considered \citep{lodato2009,guillochon2013}.

\begin{figure}
\centering
\includegraphics[width=\columnwidth]{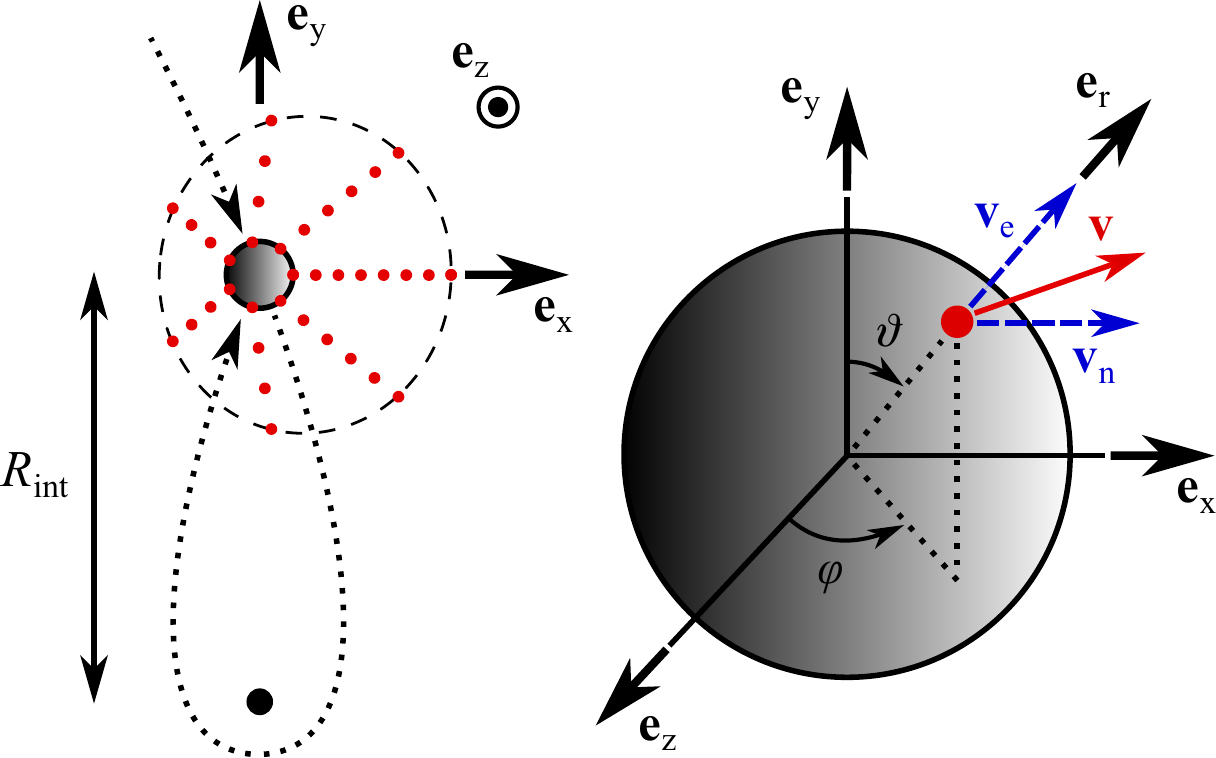}
\caption{Sketch illustrating the properties of the outflow launched from the self-crossing shock and our injection procedure used to simulate it. The left panel shows the trajectory of the stream (dotted line) around the black hole (black dot). The stream collides with itself at the intersection radius $\rint$ that drives an outflow of gas from this location. The outflow is modelled by injecting SPH particles (red points) from a spherical box displayed as the grey sphere. The frequency of injection is chosen to yield the mass outflow rate from the self-crossing shock given by equation \eqref{eq:injection-rate}. The right panel shows a zoom-in on the injection box with a single SPH particle being injected. Its position is specified by the polar angle $\vartheta$ and azimuthal angle $\varphi$ whose probability distributions are adopted to produce the density profile of equation \eqref{eq:density-profile}. The red solid arrow displays the particle velocity given by the vectorial sum of its net and expansion components $\vect{v_{\rm n}}$ and $\vect{v_{\rm e}}$ (blue dashed arrows) that are along the $\ex$ direction tangential to the black hole and the $\er$ direction pointing radially from the intersection point, respectively.}
\label{fig:sketch}
\end{figure}

When the first debris passes at pericenter, relativistic apsidal precession results in its intersection with the part of the stream still approaching the black hole. The resulting self-crossing shock causes the formation of an outflow from the intersection point. It is illustrated in the sketch of Fig. \ref{fig:sketch} along with the procedure used to model this process that we describe in the next section. The properties of the outflow are predicted from the local simulation carried out by \citet{lu2019}. For our choice of parameters, this study finds that the intersection radius is located at a distance $\rint = 381 \rg \approx 25 \rt$ from the black hole. Following the collision, the shocked gas expands spherically at a speed $\ve = 0.065 \, c$ while retaining a net velocity $\vn = 0.0152 \, c$ in the direction tangential to the black hole. In a frame co-moving at velocity $\vn \ex$, the outflow has an axisymmetric density profile with a polar dependence given by
\be
\bar{\rho}_{\rm sc}(\vartheta) = -0.08436 x^4 + 0.2434 x^3 - 0.2017 x^2 + 0.1103 x + 0.01051 
\label{eq:density-profile}
\ee
where $x \equiv \min(\vartheta,\pi-\vartheta)$ and $\vartheta$ denotes the polar angle defined in Fig. \ref{fig:sketch}. The normalization is chosen such that $2\pi\int_0^{\pi} \bar{\rho}_{\rm sc}(\vartheta) \sin \vartheta \diff \vartheta = 1$. Because the gas that passes through the self-crossing shock is immediately ejected in this quasi-spherical outflow, the corresponding total outflow rate is given by
\be
\dot{M}_{\rm sc} = \dot{M}_{\rm fb},
\label{eq:injection-rate}
\ee
where the fallback rate on the right-hand side is specified by equation \eqref{eq:fallback_rate}. Note that this prescription relies on a continuous ejection of gas from the self-crossing shock while, in reality, the mass outflow from the intersection point is intermittent. This is because the self-crossing shock is present only when provided with a stream component approaching the black hole and another going away from it after one more pericenter passage. While the former is always present, the latter gets exhausted and then replenished on the time $t_{\rm rep}$ needed for the gas element just upstream the intersection point at the moment when the self-crossing shock ends to get around the black hole and back to the intersection point. The outflow rate should then be $\dot{M}_{\rm sc} = 2 \dot{M}_{\rm fb}$ when the self-crossing shock is on and $\dot{M}_{\rm sc} = 0$ when it is off. The replenishment timescale is roughly twice the free-fall time from the intersection point, that is $t_{\rm rep} = 2 (\rint^3/G \mh)^{1/2} = 2 \days $ for our choice of parameters. The fact that $t_{\rm rep}/\tmin \approx 0.05 \ll 1$ implies that the outflow rate can be well approximated by a continuous one, justifying our treatment and the use of equation \eqref{eq:injection-rate}.

\begin{figure}
\centering
\includegraphics[width=\columnwidth]{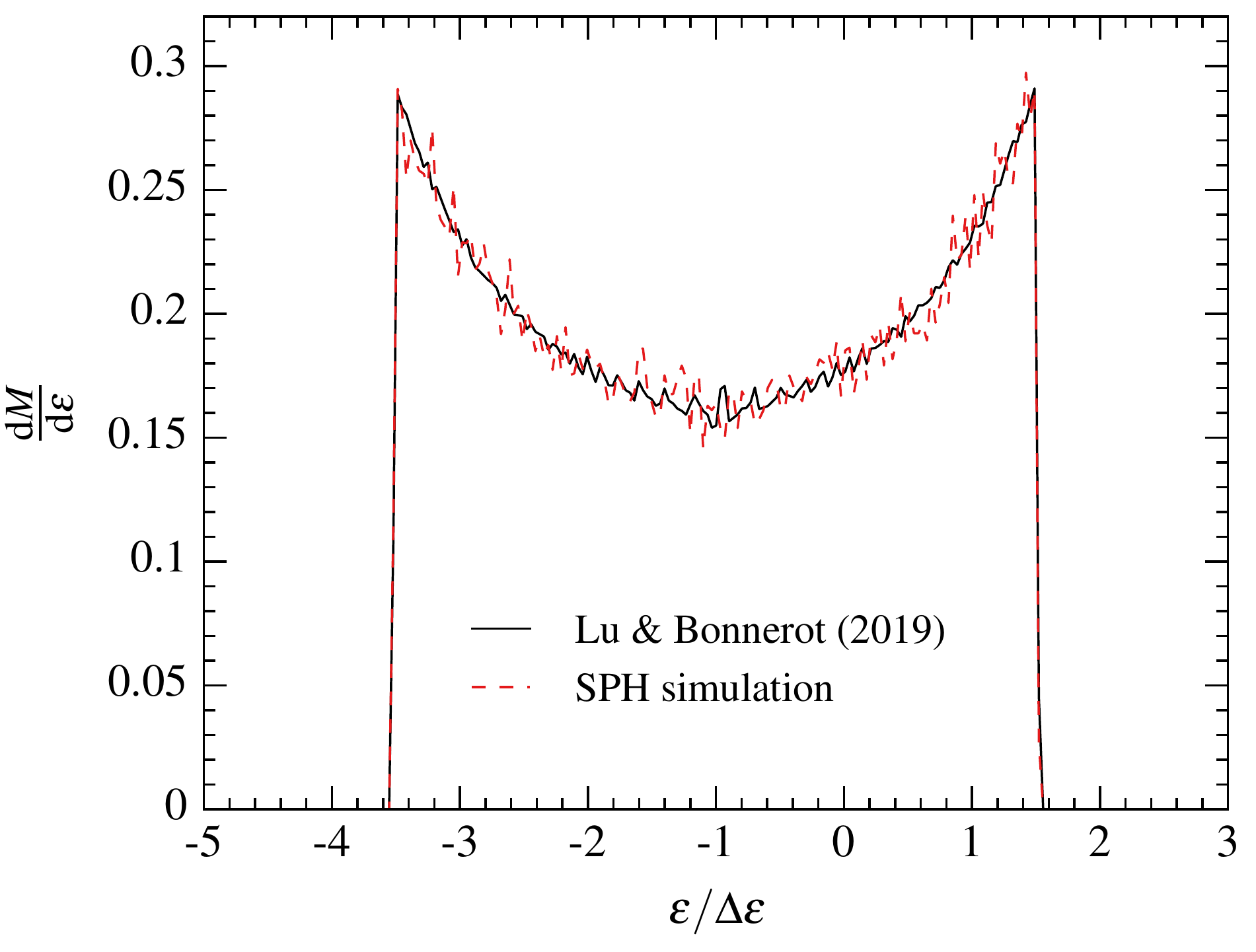}
\includegraphics[width=\columnwidth]{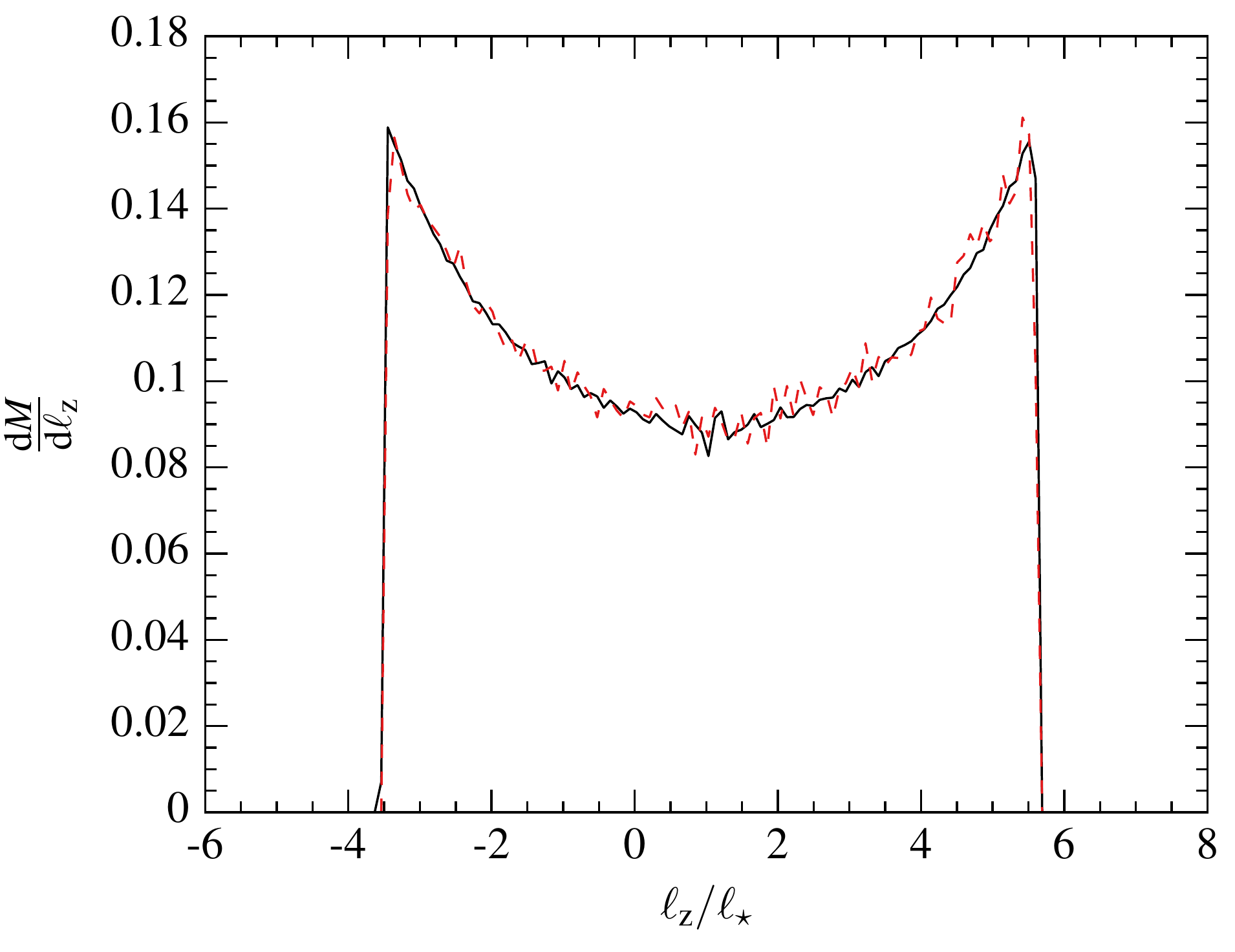}
\caption{Distributions of orbital energy (upper panel) and angular momentum projected along the $\ez$ direction perpendicular to the original stellar orbital plane (lower panel) computed from the local study of the self-crossing shock carried out by \citet{lu2019} (black solid line) and obtained directly from our SPH simulation (red dashed line) at $t/\tmin = 5 \times 10^{-3}$. The fact that the distributions obtained in these two ways match with good accuracy demonstrates the correct implementation of the injection procedure described in Section \ref{sec:numerical-setup}.}
\label{fig:initial-distributions}
\end{figure}

\subsection{Numerical setup}
\label{sec:numerical-setup}

The simulation is carried out with the SPH code \textsc{phantom} \citep{price2018} by adding gas to the computational domain according to the properties described above for the debris leaving the self-crossing shock. This procedure is illustrated in the sketch of Fig. \ref{fig:sketch}. It consists in injecting SPH particles (red dots) from a spherical box (grey sphere) of radius $\rb = 10 \rg \ll \rint$ located at the intersection point. The particles have a total velocity (red solid arrow) given by $\vect{v} = \vn \ex + \ve \er$ that is the vectorial sum of the net and expansion speeds (blue dashed arrows) directed along in the direction tangential to the black hole and radially from the injection point, respectively. Additionally, their specific thermal energy is set such that the outflow is highly supersonic.

To reproduce the outflow rate from the self-crossing shock prescribed by equation \eqref{eq:injection-rate}, a number of particles $\Delta N = \left[\mdot \Delta t / \mpart \right]$ is added to the simulation at each timestep $\Delta t$, where the brackets represent the nearest integer function.\footnote{Our simulation uses individual time-stepping, implying that SPH particles can have different timesteps. The timestep $\Delta t$ between two particle injections is the shortest among them.} The mass of the SPH particles is set to $\mpart = 10^{-8} \msun$ in the run we present in the paper. Our simulation ends at $t= \tmin$ when $37 \%$ of the bound debris have passed through the self-crossing shock, which corresponds to a total number of $N = 9.3 \times 10^6$ injected particles. The particle mass sets the resolution of the simulation with a lower $\mpart$ resulting in a higher resolution since more particles are injected during a given amount of time. In Section \ref{resolution}, we provide a resolution study by considering SPH particles of different masses and evaluate the impact on our results.

The density profile of the outflow depends on the number of particles injected at each timestep along the direction specified by the polar and azimuthal angles $\vartheta$ and $\varphi$. As described in the previous section, the desired density profile is axisymmetric, which we obtain by choosing $\varphi$ from a uniformly random distribution. The polar dependence given by equation \eqref{eq:density-profile} is instead attained by selecting $\vartheta$ according to the probability distribution $\diff N / \diff \vartheta = \bar{\rho}_{\rm sc}(\vartheta) \sin \vartheta$ where $\diff N$ is the number of particles injected with a polar angle between $\vartheta$ and $\vartheta + \diff \vartheta$. This is realized in practice by calculating the polar angle from $\vartheta = F^{-1}(s)$, where $s$ is a uniformly random number such that $0\leq s \leq 1$ and the function $F^{-1}$ is computed numerically to be the inverse of $s = F(\vartheta) \equiv 2 \pi \int_0^{\vartheta} \bar{\rho}_{\rm sc}(\vartheta') \sin \vartheta' \diff \vartheta'$ \citep{press1992}.

After it has been injected inside the computational domain, the gas evolves according to the gravity of the black hole that we model with the gravitational potential given by \citep{tejeda2013}
\be
\Phi = -\frac{G \mh}{R} - \left( \frac{2 \rg}{R- 2 \rg} \right) \left[ \left( \frac{R- \rg}{R- 2 \rg} \right) \dot{R}^2 + \frac{ R^2 (\dot{\theta}^2 + \sin^2 \theta \, \dot{\phi}^2)}{2} \right], 
\label{eq:potential}
\ee
where $R$ is the distance from the black hole and dotted variables are derived with respect to time. Here, $\theta$ and $\phi$ denote the polar and azimuthal angles of a test particle in spherical geometry. Note that they are different from those $\vartheta$ and $\varphi$ defined earlier that denote the location on the injection box. This potential reproduces many features of the Schwarschild metric including relativistic apsidal precession. It has been been implemented in \textsc{phantom} and tested by \citet{bonnerot2016-circ}. We also ran the same simulation adopting a Keplerian potential and found no significant differences. A test particle evolving under the gravitational potential of equation \eqref{eq:potential} conserves the specific orbital energy and angular momentum given by 
\be
\epsilon = \frac{1}{2}\left[\frac{R^2 \dot{R}^2}{(R - 2 \rg)^2} + \frac{R^3 (\dot{\theta}^2 + \sin^2 \theta \, \dot{\phi}^2) }{R - 2 \rg}  \right] - \frac{G \mh}{R},
\label{eq:energy}
\ee
\be
\ell = \frac{R^3 (\dot{\theta}^2 + \sin^2 \theta \, \dot{\phi}^2)^{1/2}}{R - 2 \rg},
\label{eq:angmom}
\ee
respectively \citep{tejeda2013}. As expected, these two quantities converge to the Keplerian values in the limit $R \gg \rg$.

We assume that the gas evolves adiabatically away from shocks where non-adiabatic heating occurs. Because pressure is provided by radiation, we adopt an adiabatic exponent of $\gamma = 4/3$. To treat shocks, we rely on a standard implementation of artificial viscosity. In combination, we make use the switch designed by \citet{cullen2010} that accurately detects shocks and only switches on artificial viscosity at their location. The assumption of adiabaticity is due to the optical thickness of the gas that prevents it from radiatively cooling. We justify this assumption in Section \ref{sec:radiation} where we carry out post-processing analyses to evaluate the impact of photon diffusion. Self-gravity is insignificant for this problem because it is overwhelmed by pressure gradients due to heating by shocks and we therefore do not consider it. Accretion onto the compact object is modelled by an accretion radius located at $R_{\rm acc} = 6\rg$ that is the innermost stable circular orbit for a non-rotating black hole. When an SPH particle enters this radius, it is removed from the simulation.

\subsection{Early dynamics of the injected gas}

\label{sec:early-dynamics}

Before fully presenting our results, it is worth emphasizing the early dynamics of the gas under the gravity of the black hole after it leaves the self-crossing shock according to the outflow properties described in Section \ref{sec:initial-conditions} for our choice of parameters. Before the self-crossing, the fluid elements within the stream share the same energy and angular momentum given by the energy spread $-\Delta \epsilon$ and the stellar angular momentum $\ell_{\star}$, respectively. The outflow launched from the intersection point leads to a redistribution of these orbital elements around these values.

\begin{figure*}
\centering
\vspace{0.6cm}
\includegraphics[width=\textwidth]{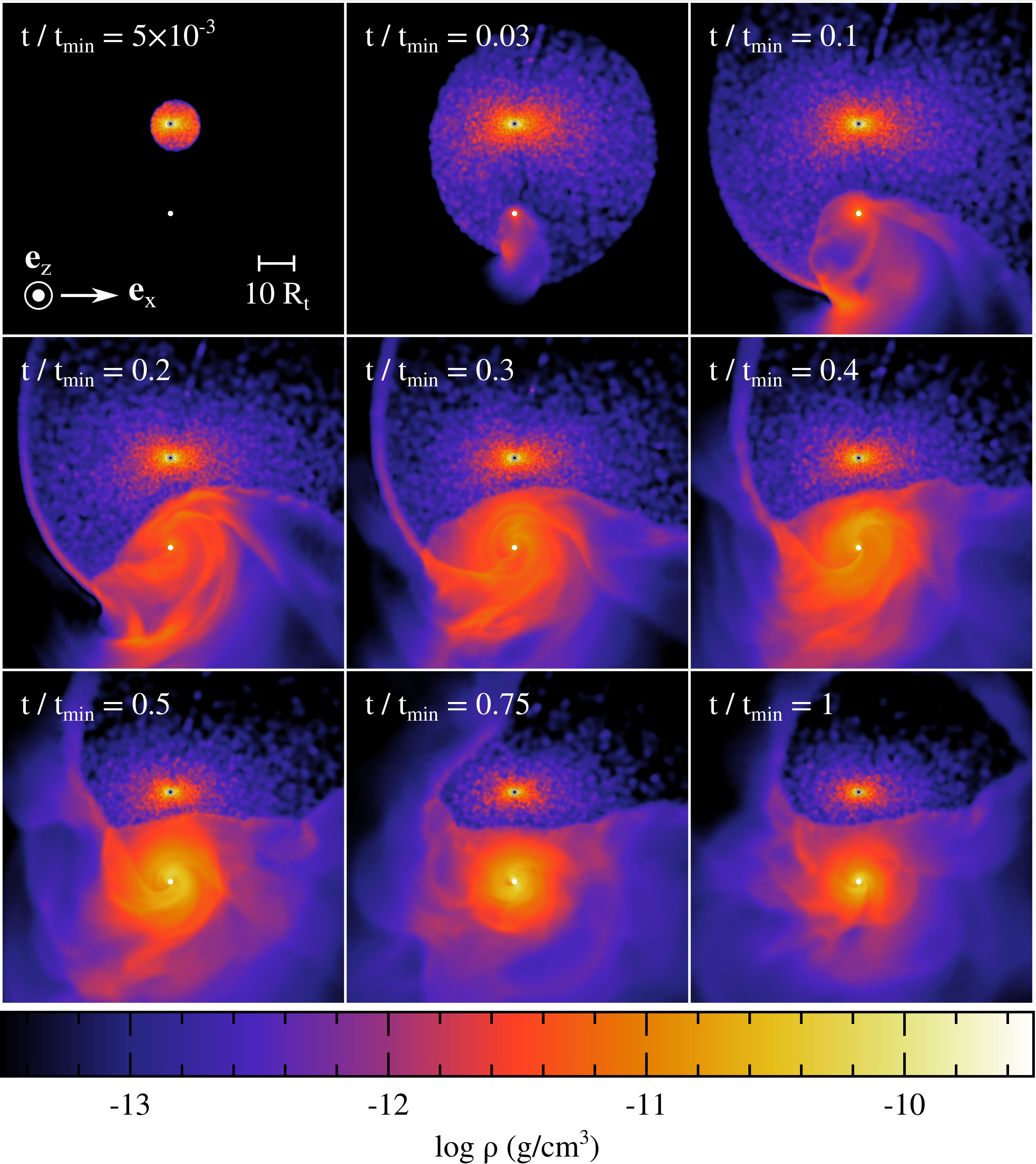}
\caption{Snapshots showing the gas distribution at different times $t/\tmin =5\times 10^{-3}$, 0.03, 0.1, 0.2, 0.3, 0.4, 0.5, 0.75 and 1 in a slice parallel to the orbital plane and containing the black hole. The colours represent the density increasing from black to white as indicated on the color bar. The black hole is represented by the white dot while the small grey circle indicates the intersection point, from which matter is injected in the computational domain. The white segment on the first panel shows the scale while the two vectors indicate the $\ex$ and $\ez$ directions, the latter being perpendicular to the equatorial plane.}
\label{fig:density-time-xy}
\end{figure*}

We display in Fig. \ref{fig:initial-distributions} the corresponding distributions of orbital energy (upper panel) and angular momentum projected along the $\ez$ direction orthogonal to the original stellar orbital plane (lower panel) in the outflowing gas normalized such that the area below the curves is one. They are computed both from the local simulation of \citet{lu2019} (black solid line, also shown with the green histogram in their figure 8) and directly from our simulation (red dashed line) at $t/\tmin = 5 \times 10^{-3}$. As can be seen from the upper left snapshot of Fig. \ref{fig:density-time-xy}, the gas is still close to the injection point at this time and has not yet experienced significant interactions such that each fluid element has conserved their energy and angular momentum from the moment of injection. The fact that the distributions obtained from these two methods match to a good accuracy demonstrates that our injection procedure is correctly implemented.

The early motion of the gas is entirely specified by the distributions of Fig. \ref{fig:initial-distributions}. In particular, it can be seen that a significant fraction around $33\%$ of the shocked gas has a positive orbital energy $\epsilon >0$ and is therefore unbound. This unbound component corresponds to matter being accelerated in the positive $\ex$ direction of its net speed during the outflow that results in a total velocity large enough to escape the black hole \citep{lu2019}. Additionally, some of the gas has a negative angular momentum $\ell_{\rm z}<0$, meaning that it rotates in the direction opposite to that of the star. It corresponds to matter launched near the $-\ex$ direction opposite to its net speed. Although most of the gas leaving the self-crossing shock is still co-rotating, the majority moves along the $\ex$ direction such that it gets unbound and leaves the system. As a result, the bound gas has on average a negative angular momentum and orbits the black hole in the counter-rotating direction. These initial orbital properties are determinant for the subsequent gas evolution studied in the rest of the paper.

\begin{figure*}
\centering
\vspace{0.6cm}
\includegraphics[width=\textwidth]{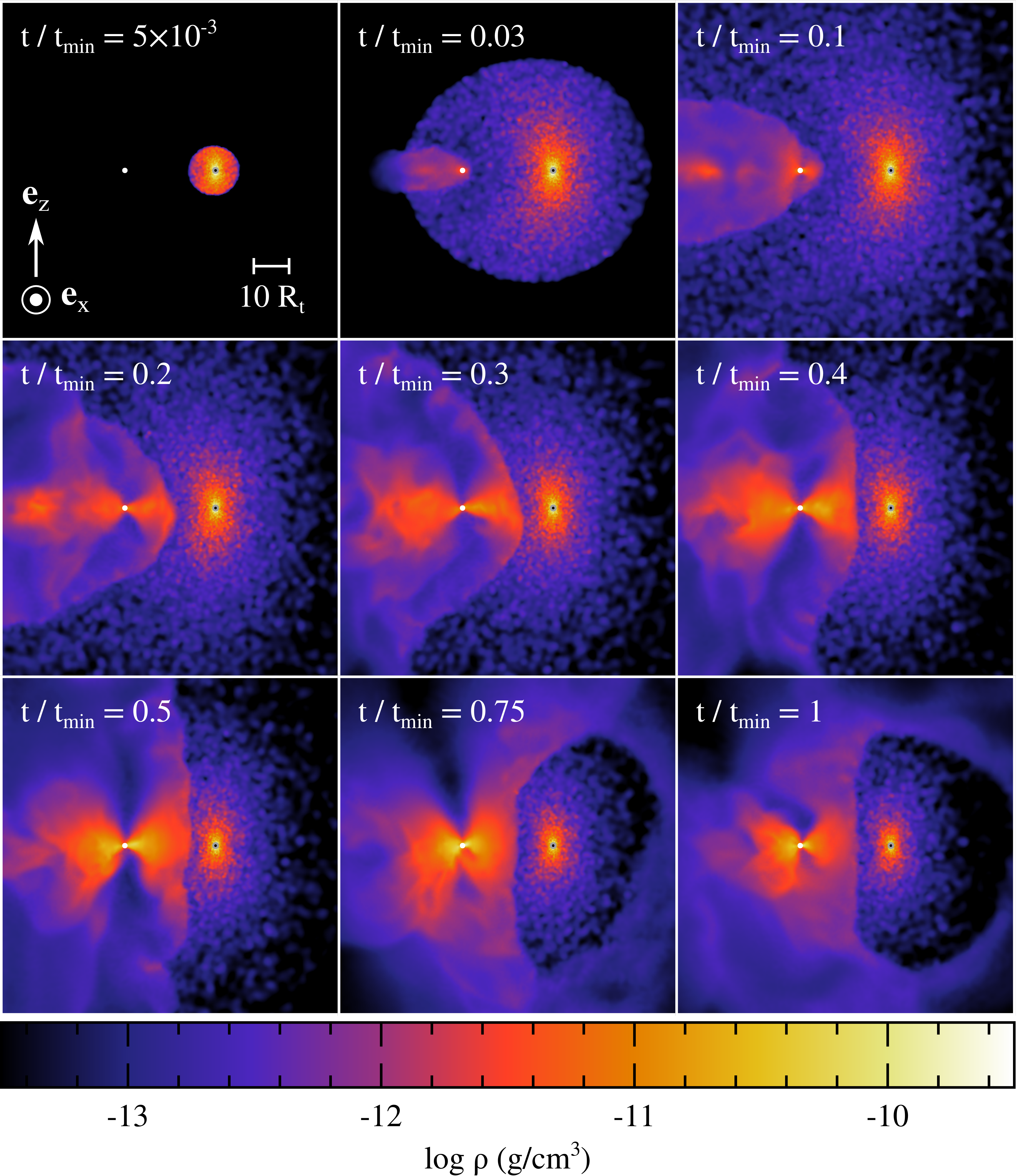}
\caption{Snapshots showing the gas distribution at different times $t/\tmin =5\times 10^{-3}$, 0.03, 0.1, 0.2, 0.3, 0.4, 0.5, 0.75 and 1 in a slice orthogonal to the orbital plane and containing the black hole. The colours represent the density increasing from black to white as indicated on the color bar. The black hole is represented by the white dot while the small grey circle denotes the intersection point, from which matter is injected in the computational domain.  The white segment on the first panel shows the scale while the two vectors indicate the $\ex$ and $\ez$ directions, the latter being perpendicular to the equatorial plane.}
\label{fig:density-time-yz}
\end{figure*}

\begin{figure}
\centering
\includegraphics[width=\columnwidth]{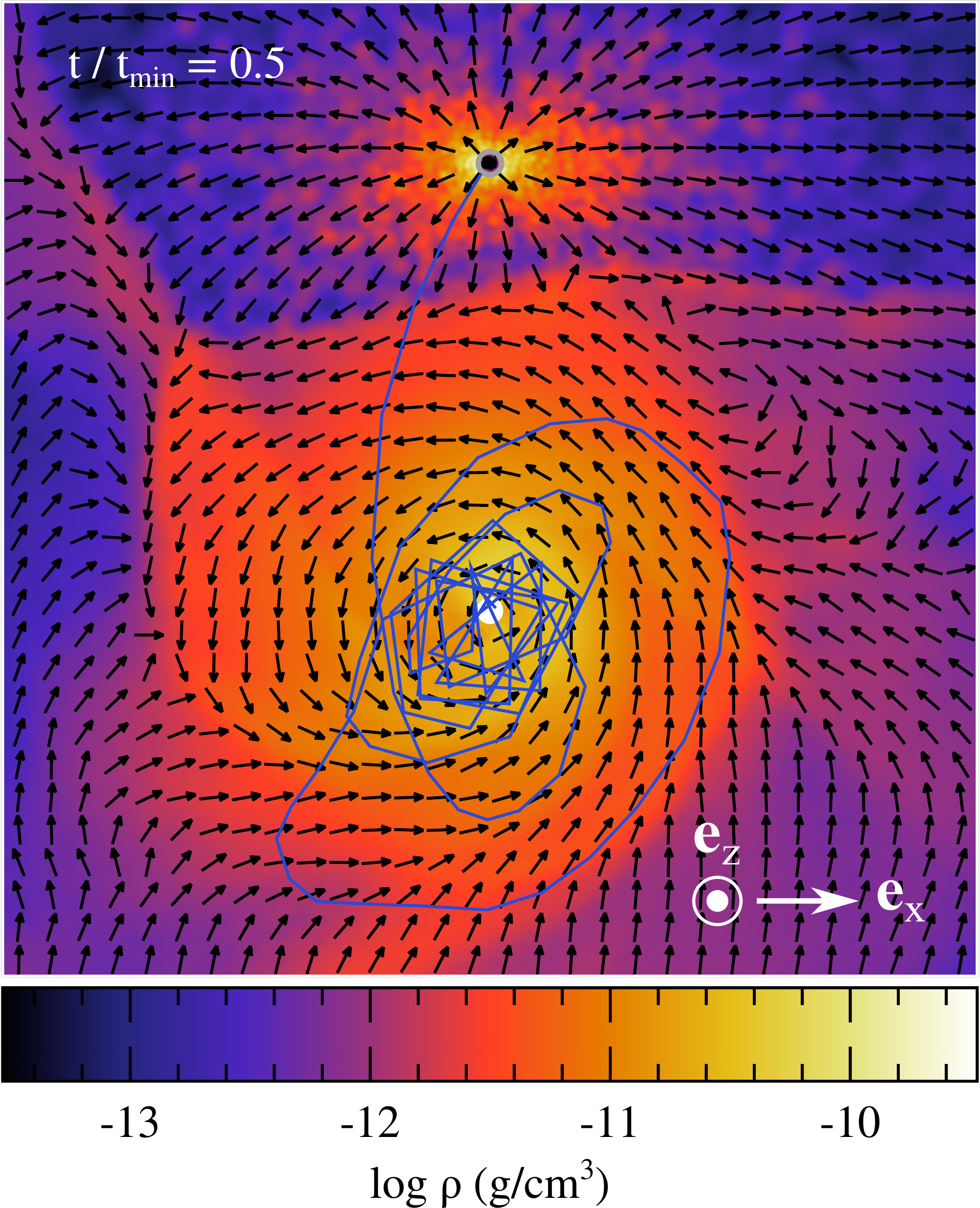}
\caption{Gas density in a slice parallel to the orbital plane of the original star at a time $t/\tmin = 0.5$. The direction of the velocity field is indicated by the black arrows that all have the same length. The blue overplotted line represents the historic trajectory of an SPH particle. The grey circle indicates the location of the injection point while the white dot represents the black hole.}
\label{fig:trajectories-xy}
\end{figure}

\begin{figure}
\centering
\includegraphics[width=\columnwidth]{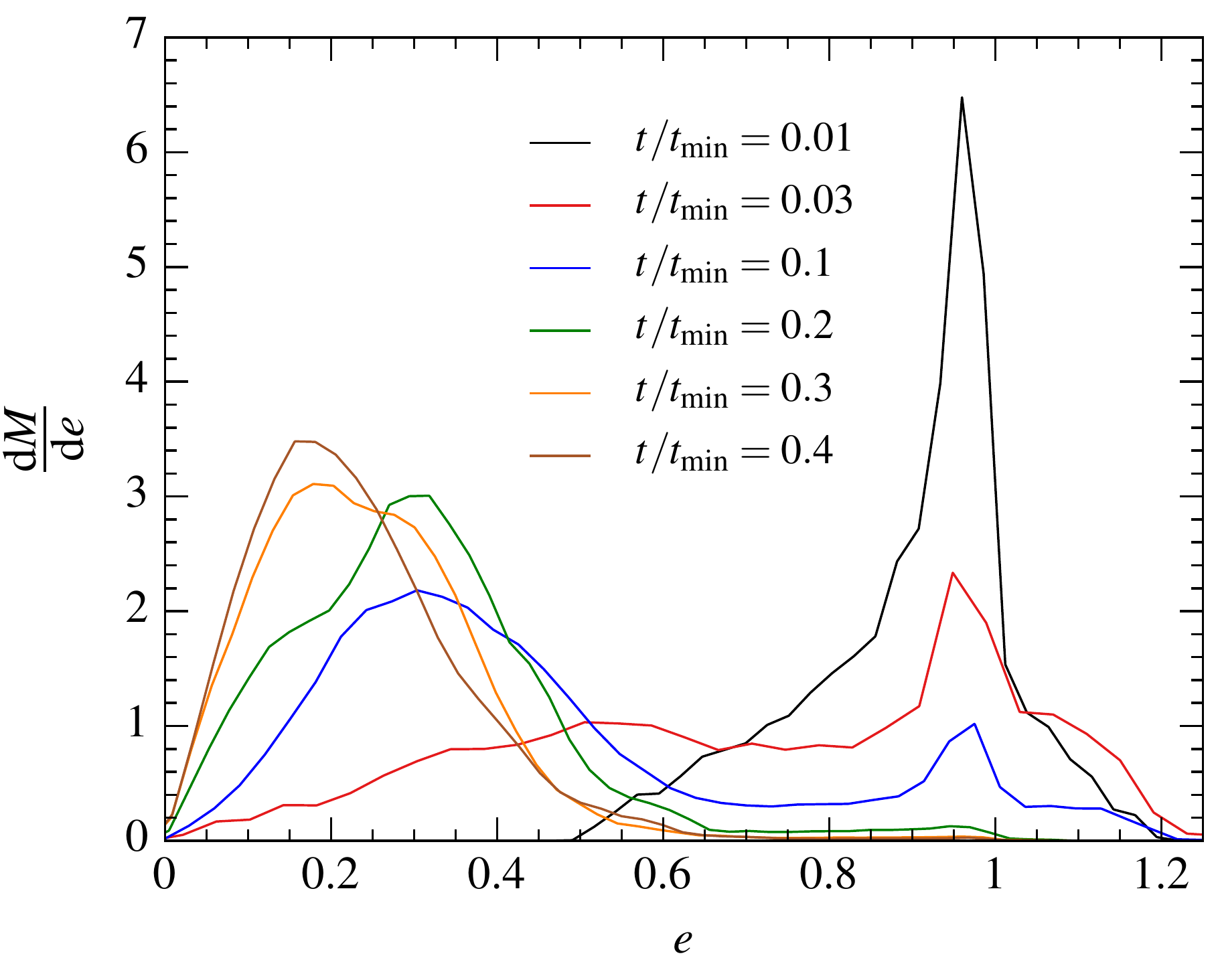}
\caption{Eccentricity distribution of the stellar debris at different times $t/\tmin = 0.01$ (black line), 0.03 (blue line), 0.1 (red line), 0.2 (green line), 0.3 (orange line) and 0.4 (brown line) for the gas inside the forming disc located at a distance $R \leq 20 \rt$ from the black hole and within a polar angle $\pi/8$ from the equatorial plane. The distribution is normalized such that the area below each curve is one.}
\label{fig:eccentricities}
\end{figure}

\section{Results}
\label{sec:results}

We now present the results of the simulation to determine the fate of the gas injected from the intersection point by the self-crossing shock.\footnote{Movies of the simulation presented in this paper are available at \url{http://www.tapir.caltech.edu/~bonnerot/realistic-disc.html}.} The simulation is carried out using the same conventions as displayed in the sketch of Fig. \ref{fig:sketch} with the star and stream of debris rotating in the clockwise direction. The origin of the coordinate system is set at the location of the black hole and $t  =0$ corresponds to the moment when the first fluid element is injected into the computational domain.

\subsection{Disc assembly}

\subsubsection{Secondary shocks}
\label{sec:secondary}

After being injected into the computational domain from the self-crossing point, the gas moves on a wide range of trajectories that are prone to crossing each others, leading to secondary shocks. The resulting gas evolution is displayed in Figs. \ref{fig:density-time-xy} and \ref{fig:density-time-yz} that show its density at different times in slices parallel to the orbital plane of the original star and orthogonal to it, respectively. The first interaction happens at $t/\tmin \approx 0.03$, involving bound matter with the lowest angular momentum promptly reaching pericenter with opposite directions of rotation. The trajectories of this gas converge on the side of the black hole opposite to the intersection point leading to almost head-on collisions. The shocked region is initially confined near the black hole but rapidly expands in all directions. By $t/\tmin \approx 0.3$, this process leads to the appearance of a clearly defined boundary between the cold infalling matter and the heated gas located at $y \approx 15 \rt$ that is approximately halfway between the black hole and the injection point. In addition, interactions take place at smaller distances that cause more dissipation owing to the larger relative velocities involved.

Furthermore, this gas distribution displays an overdense filament inside the orbital plane of the original star, most visible in Fig. \ref{fig:density-time-xy} on the left-hand side of the snapshots corresponding to times $0.1 \leq t/\tmin \leq 0.4$. It is due to weak shocks taking place at radii $R \gtrsim 20 \rt$ that involve bound gas getting away from the self-crossing point and debris injected earlier that approaches the compact object after having reached its apocenter. The gas making up this filament continuously approaches the black hole and perturbs the rotating structure forming at lower distances. Remarkably, the associated overdensity seems to extend deeply into the accretion disc, as is most evident at $t/\tmin = 0.4$. We attribute this phenomenon to the formation of spiral shocks in the forming gaseous disc due to the external perturbation produced by the returning debris, as we further analyse in Section \ref{sec:spiral-shocks}.

\subsubsection{Circularization in a counter-rotating disc}

The numerous secondary shocks forming as the gas reaches the black hole lead to kinetic energy dissipation. As a result, the gas progressively assumes more circular orbits to form a coherent rotating structure that settles around $t/\tmin \approx 0.3$. Remarkably, because most of the injected bound gas rotates in the direction opposite to that of the original star as explained in Section \ref{sec:early-dynamics}, the resulting disc is also counter-rotating. This can be seen from Fig. \ref{fig:trajectories-xy} that shows the gas density at $t/\tmin = 0.5$ in the newly-formed disc overplotted with black arrows of the same size that denote the direction of the gas velocity field. While the star's trajectory was in the clockwise sense, the disc forming from the debris turns in the opposite, counter-clockwise way.

As can be seen from Fig. \ref{fig:density-time-xy}, this gaseous disc extends to a distance of $R \approx 15 \, \rt = 0.2 \,\amin$. This property results from the spread in angular momentum imparted by the self-crossing shock as well as secondary shocks. It confirms that the gas does not settle into a circular ring at $R = 2 \rt$. This latter assumption is commonly made in the literature but is only expected if angular momentum exchange is negligible from the stellar disruption onwards, which we find is not the case.\footnote{If all the fluid elements were able to circularize while keeping a common angular momentum equal to that of the star $\ell_{\star} = (2 G \mh \rt)^{1/2}$, the resulting disc would be a ring located at the circularization radius $R = \lstar^2/(G \mh) = 2 \rt$. Because angular momentum gets efficiently redistributed among the debris through multiple shocks, we find that the trajectories of the gas within the disc span instead a large range of radii.} Furthermore, Fig. \ref{fig:density-time-yz} shows that the newly-formed disc is very thick with an aspect ratio $H/R \approx 1$ as a consequence of the thermal energy injection imparted by the circularizing shocks. It also displays two clearly defined funnels where the gas density is lower than along the disc mid-plane by up to two orders of magnitude.

\begin{figure}
\centering
\includegraphics[width=\columnwidth]{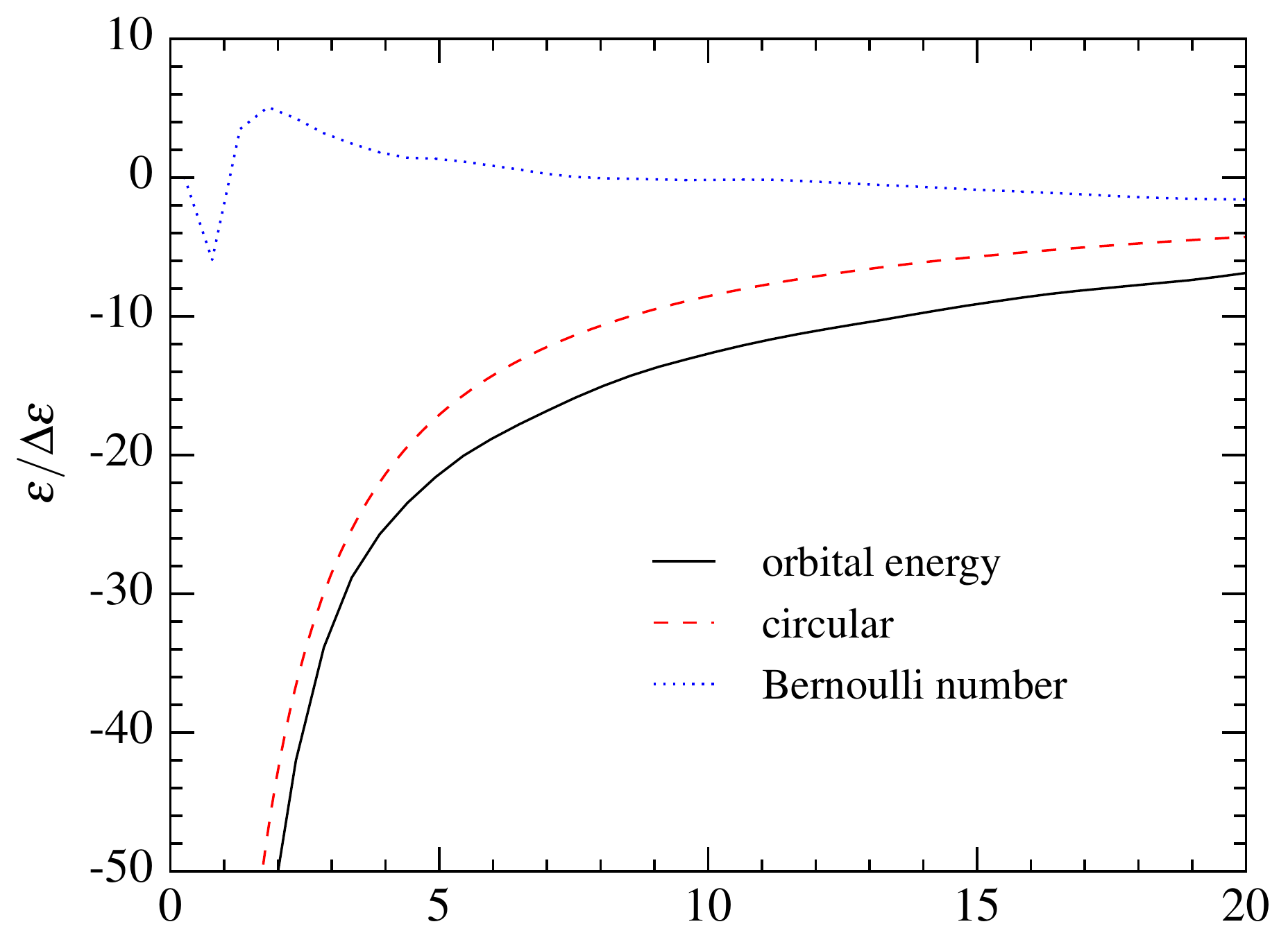}
\includegraphics[width=\columnwidth]{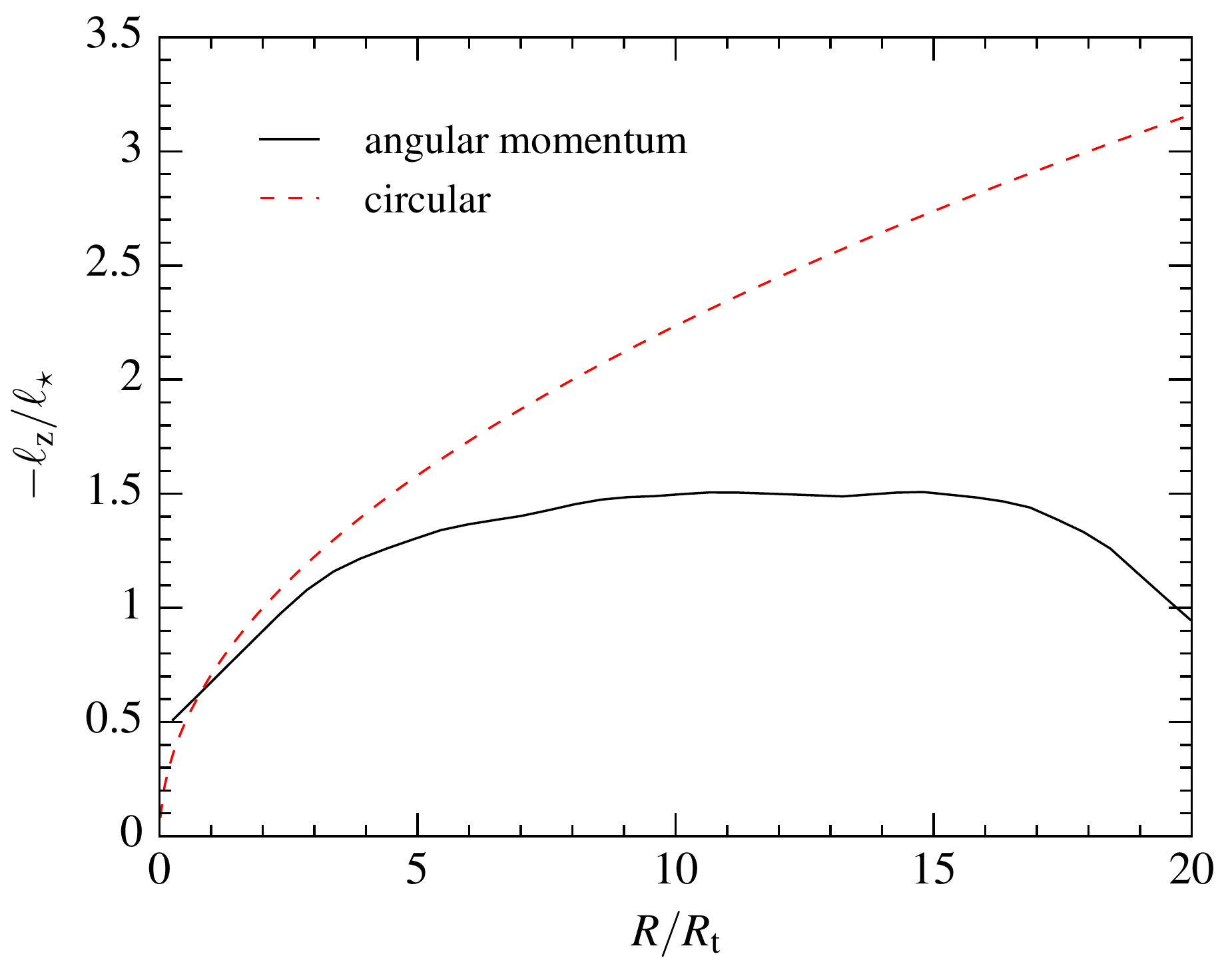}
\caption{Spherically-averaged profiles of energy (upper panel) and angular momentum projected along the $\ez$ direction (lower panel) at $t/\tmin = 0.5$ for the gas located within a polar angle $\pi/8$ from the equatorial plane (black solid line) and compared to the same orbital elements assuming circular motion in a Keplerian potential given by $\epsilon_{\rm c} = -G \mh/(2 R)$ and $\ell_{\rm c} = (G \mh R)^{1/2}$ (red dashed line). The blue dotted line in the upper panel denotes the Bernoulli number of the gas obtained from $\mathcal{B} = \epsilon + 4 P/\rho$.}
\label{fig:energy-angmom}
\end{figure}

Gas circularization can be evaluated more quantitatively from the evolution of the eccentricity distribution shown in Fig. \ref{fig:eccentricities} at different times and considering the debris located at a distance $R \leq 20 \rt$ from the black hole and within a polar angle $\pi/8$ from the equatorial plane, that is such that $|\theta - \pi/2| \leq \pi/8$. This region contains an increasing mass of gas but, for clarity, all the distributions are normalized such that the area below them are equal to one. Although the gas trajectories are sub-Keplerian because of the large pressure gradients pointing outward, we define an effective eccentricity that approximately describes the shape of the orbits by replacing the azimuthal component of the velocity $v_{\phi}$ by the larger value given by $v'_{\phi} = (v^2_{\phi}+ R| \nabla P/\rho|)^{1/2}$, where the pressure gradient is estimated by $R| \nabla P/\rho|\approx 8 P / 3 \rho$ using the fact that the pressure scales as $P \propto \rho^{4/3} \propto R^{-8/3}$ since the gas evolves adiabatically and the density decreases roughly as $R^{-2}$ in most of the domain considered (see Section \ref{sec:outflow}). From this increased azimuthal component, the eccentricity is then computed according to Keplerian dynamics through $e = \sqrt{1 + 2\epsilon \ell^2 / (G \mh)^2}$, which is valid since most of the matter is located at distances from the black hole much larger than $\rg$. At $t/\tmin = 0.01$, none of the gas has circularized yet since it has not experienced any interaction. The peak around $e \approx 1$ is present because only the most eccentric gas can reach small radii within this time. We emphasize that it is not representative of the whole distribution of injected bound gas whose eccentricities are closer to $e \approx 0.5$. The distribution widens around $t/\tmin = 0.03$ as more gas can make it in the region considered and the secondary shocks cause some of the orbits to become close to circular. As time goes on, the eccentricity distribution shifts to lower values while remaining significantly broad. It settles down around $t/\tmin =0.3$ that corresponds to the completion of disc formation according to the gas evolution presented above. The final distribution peaks around $e = 0.2$ with a large spread around this value.

The gas retains non-negligible eccentricities as a result of highly irregular motion even after the debris has settled into a coherent rotating structure. This point is illustrated by the overplotted blue line in Fig. \ref{fig:trajectories-xy} that represents a typical trajectory for the gas inside the disc obtained by following a single SPH particle of our simulation. After joining the disc from the injection point, this gas element experiences large and violent variations in its orbital elements such that its trajectory covers a wide range of locations both within the mid-plane and away from it. Note that the line has sharp corners when the fluid element reaches small radii. This feature is artificial and results from the fact that this trajectory is recorded for time intervals larger than the orbital timescale of the particle. This sporadic motion of the gas inside the disc is also accompanied by variations of its angular momentum vector whose direction randomly changes that can result in a misalignment of up to a few degrees with the $\ez$ direction orthogonal to the stellar orbital plane.

\subsubsection{Overall disc dynamics}
\label{sec:overall}

The global dynamics of the disc is analysed more in details in Fig. \ref{fig:energy-angmom} that shows with solid black lines the azimuthally-averaged profiles of specific orbital energy (upper panel) and angular momentum projected along the original stellar orbital plane (lower panel) obtained from equation \eqref{eq:energy} and \eqref{eq:angmom} at $t/\tmin = 0.5$ and considering the gas located within a polar angle $\pi/8$ from the equatorial plane. Note that the gas angular momentum is opposite to that of the star with $\ell_{\rm z}/\ell_{\star}<0$ because the disc is counter-rotating as explained above. The red lines denote the values of the orbital energy and angular momentum assuming circular motion in a Keplerian potential given by $\epsilon_{\rm c} = -G \mh/(2 R)$ and $\ell_{\rm c} = (G \mh R)^{1/2}$, respectively. They are larger than that of the gas at all distances that is a consequence of both its eccentric trajectories and the fact that the disc is supported by pressure gradients in addition to the centrifugal force. For $R\lesssim 5\rt$, the gas angular momentum nevertheless approaches its circular value with $\ell_{\rm z} \approx \ell_{\rm c}$, implying that centrifugal support dominates in this inner region. At larger distances, this profile is nearly flat with an inversion around $R\approx 20\rt$ due to the contribution from the low angular momentum gas directly coming from the injection point on almost radial trajectories.

The blue dotted line in the upper panel of Fig. \ref{fig:energy-angmom} shows the Bernoulli number of the gas given by $\mathcal{B} = \epsilon + 4 P/\rho$ for the adiabatic exponent $\gamma = 4/3$. The second term takes into account the energy increase resulting from heating by secondary shocks. This number is close to zero throughout the disc, slowly declining outwards. The decrease at $R \approx \rt$ is caused by the pressure drop associated to the removal of the gas crossing the accretion radius. At small radii $R \lesssim 5 \rt$, it becomes positive due to the large internal energy injection happening in this region. The fact that $\mathcal{B} \approx 0 \gg \epsilon_{\rm c}$ for most of the gas implies that it can reach large distances or even get unbound from the system if its heat is converted into kinetic energy. This situation is prone to strong gas outflow that will be further examined in Section \ref{sec:outflow}. The marginally-bound nature of this gaseous structure is also consistent with analytical models of TDE discs, such as the one developed by \citet{coughlin2014}.

\begin{figure}
\centering
\includegraphics[width=\columnwidth]{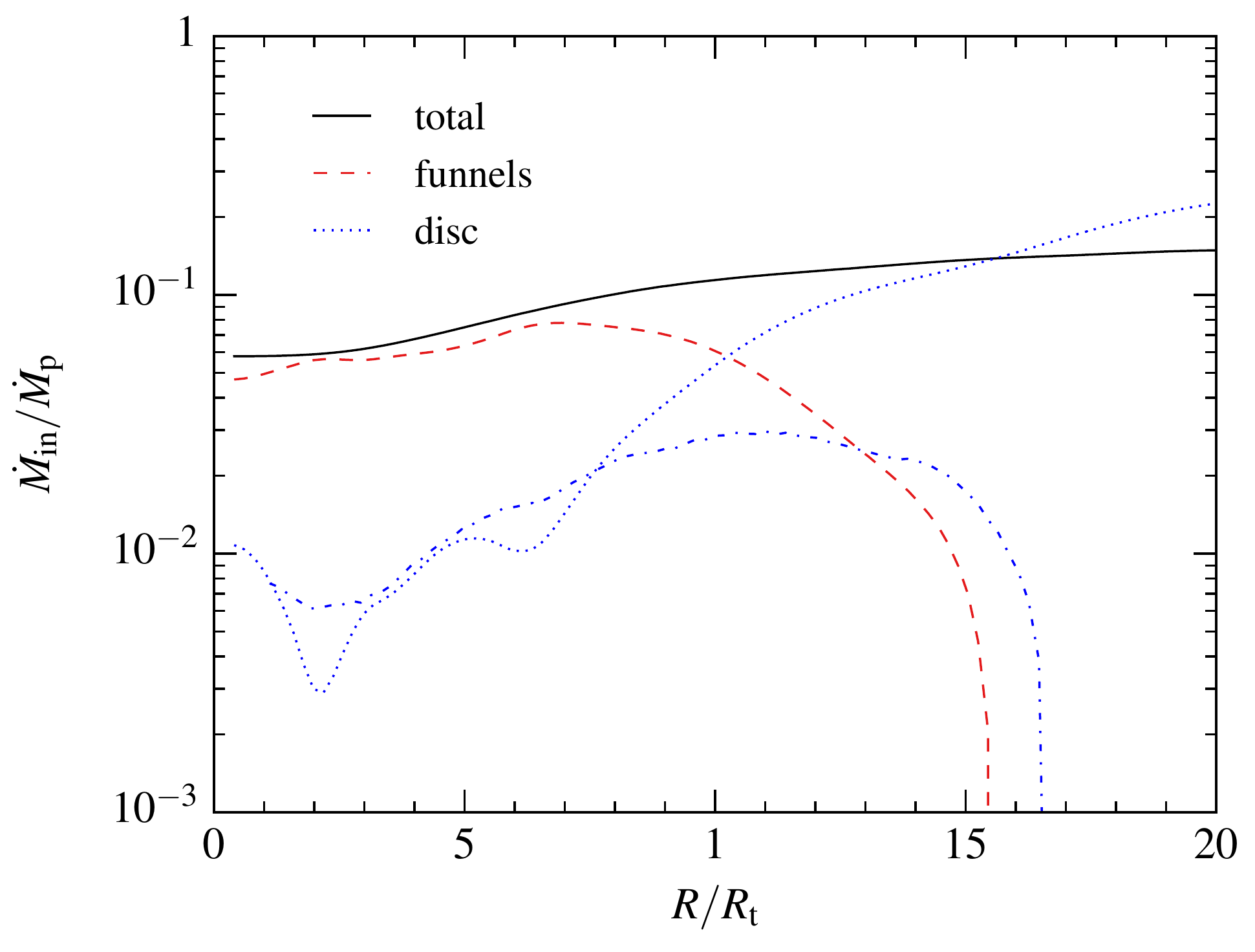}
\caption{Radial profile of the total net inflow rate (black solid line) time-averaged on the interval $0.4 \leq t/\tmin \leq 0.5$ and the same calculation considering only the gas originating from the funnels (dashed red line) and the disc (blue dotted line). The disc is defined as the region within a polar angle $\pi/8$ from the equatorial plane while the funnels occupy the rest of the space. The dashed-dotted blue line shows the mass inflow rate due to spiral shocks obtained from the induced stress according to equation \eqref{eq:mdotst}.}
\label{fig:mdotinvsr}
\end{figure}

\begin{figure}
\centering
\includegraphics[width=\columnwidth]{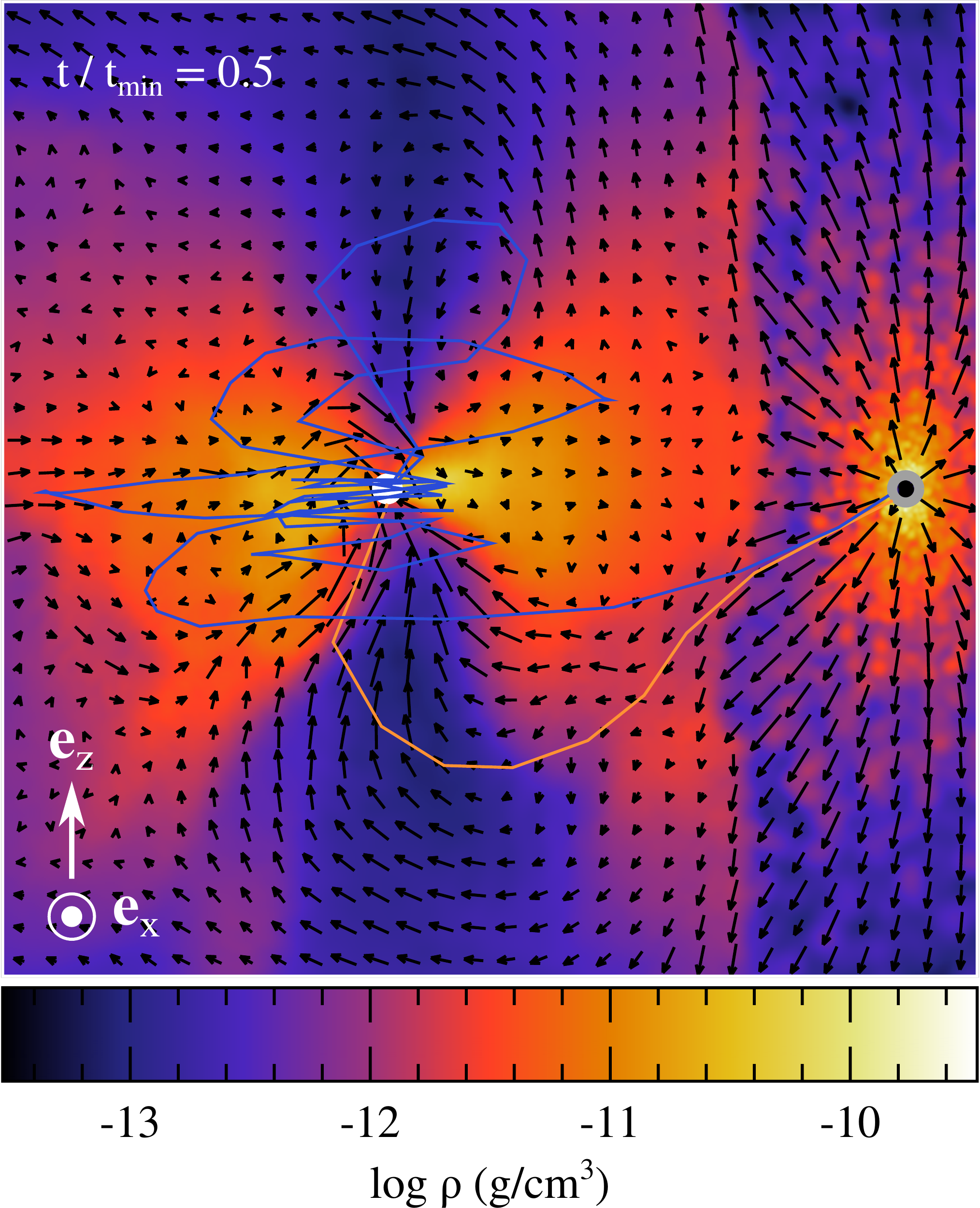}
\caption{Gas density in a slice orthogonal to the orbital plane of the original star. The black arrows represent both the magnitude and direction of the velocity field. The blue and orange overplotted lines indicate the trajectories of SPH particles that end up crossing the accretion radius after falling along the funnel of the disc. The grey circle indicates the location of the injection point while the white dot represents the black hole.}
\label{fig:trajectories-zy}
\end{figure}

\subsection{Inflow and accretion}
\label{sec:inflow-accretion}

\subsubsection{Gas inflow}
\label{sec:inflow}

The gas present in the inner regions moves on average towards the black hole. We compute the rate at which this matter flows inward from $\dot{M}_{\rm in} = (\Delta M_{\rm in} - \Delta M_{\rm out})/\Delta t$ where $\Delta M_{\rm in}$ and $\Delta M_{\rm out}$ denote the masses of gas that cross a sphere of given radius during a time-step $\Delta t$ going inward and outward, respectively. The radial profile of this total net inflow rate time-averaged on the interval $0.4 \leq t/\tmin \leq 0.5$ is displayed in Fig. \ref{fig:mdotinvsr} (black line) that also shows the same calculation considering only the gas originating from the funnels (red dashed line) and the disc (blue dotted line). As before, gas is defined to belong to the disc if it is located within a polar angle $\pi/8$ from the equatorial plane while the rest of the matter is part of the funnels. The total inflow rate  increases only slowly with radius with a value $\dot{M}_{\rm in} \lesssim 0.1 \, \mdotp \approx  2.5  \, \mdotedd$. However, its origin changes drastically with distance from the black hole. At $R \gtrsim 10 \rt$, most of the inward motion occurs close to the mid-plane while the gas inside the funnels moves either slower inward or outward. This inflow is mostly caused by matter joining the outer edges of the disc either coming directly from the injection point or from larger distances after being previously expelled by secondary shocks. This hierarchy inverses closer in where inflow along the funnels dominates that inside the disc by about an order of magnitude, the latter decreasing to $\sim 10^{-2} \mdotp \approx 0.25  \, \mdotedd$. We highlight that this difference cannot be accounted for by the different solid angles covered by the two regions since that of the funnels is larger by a factor of only $(1-\sin (\pi/8))/\sin (\pi/8) \approx 1.6$. The larger inflow rate along the funnels in the inner region is therefore caused by a higher mass flux. This polar angle dependence of the inflowing mass flux has also been found in the simulation of disc formation by \citet{sadowski2016}.

The main characteristics of the gas radial motion can be seen directly from Fig. \ref{fig:trajectories-zy} that shows the gas density in a slice orthogonal to the equatorial plane at $t/\tmin=0.5$ with black arrows of different lengths indicating both direction and magnitude of the velocity field. While there is hardly any inward motion visible inside the disc, the matter contained in the funnels is clearly falling towards the black hole that is possible due to lower pressure and centrifugal support in these regions. The blue line indicates the trajectory of a given SPH particle that is representative of the inflowing gas. After being injected from the intersection point, this fluid element joins the disc and revolves several times around the black hole. However, it eventually gets ejected from the mid-plane to the upper funnel, along which it quickly falls inward all the way inside the accretion radius. The orange line shows the trajectory of another gas element that enters the lower funnel and gets accreted in one single infall after being injected from the self-crossing point with only a mild change in its orbital elements. We find that this second possibility is, however, less likely with most of the mass inside the funnels being injected at least a few dynamical timescales earlier.

\begin{figure}
\centering
\includegraphics[width=\columnwidth]{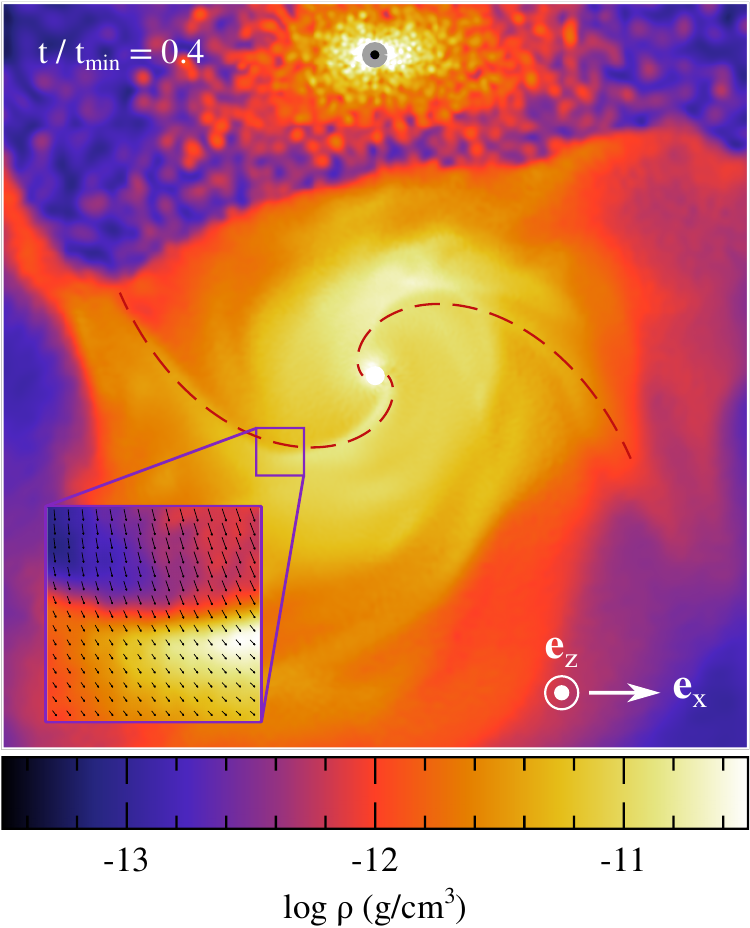}
\caption{Snapshot showing the gas density in a slice parallel to the equatorial plane at $t/\tmin = 0.4$. The grey circle indicates the location of the injection point while the white dot represents the black hole. The red dashed line displays a fit to the two overdense features using equation \eqref{eq:spiral} derived from the dispersion relation for spiral shocks with $m = 2$, $\mathcal{M} = 1.3$, $R_0 = \rt$ and $\phi_0 = 0$. The inset shows a zoom-in on a part of the leftmost spiral with colours showing the gas density rescaled to make the feature more apparent with an increase of a factor of only $\sim 2$ from blue to orange. The black arrows show both the direction and magnitude of the velocity field.}
\label{fig:dispersion}
\end{figure}

\begin{figure}
\centering
\includegraphics[width=\columnwidth]{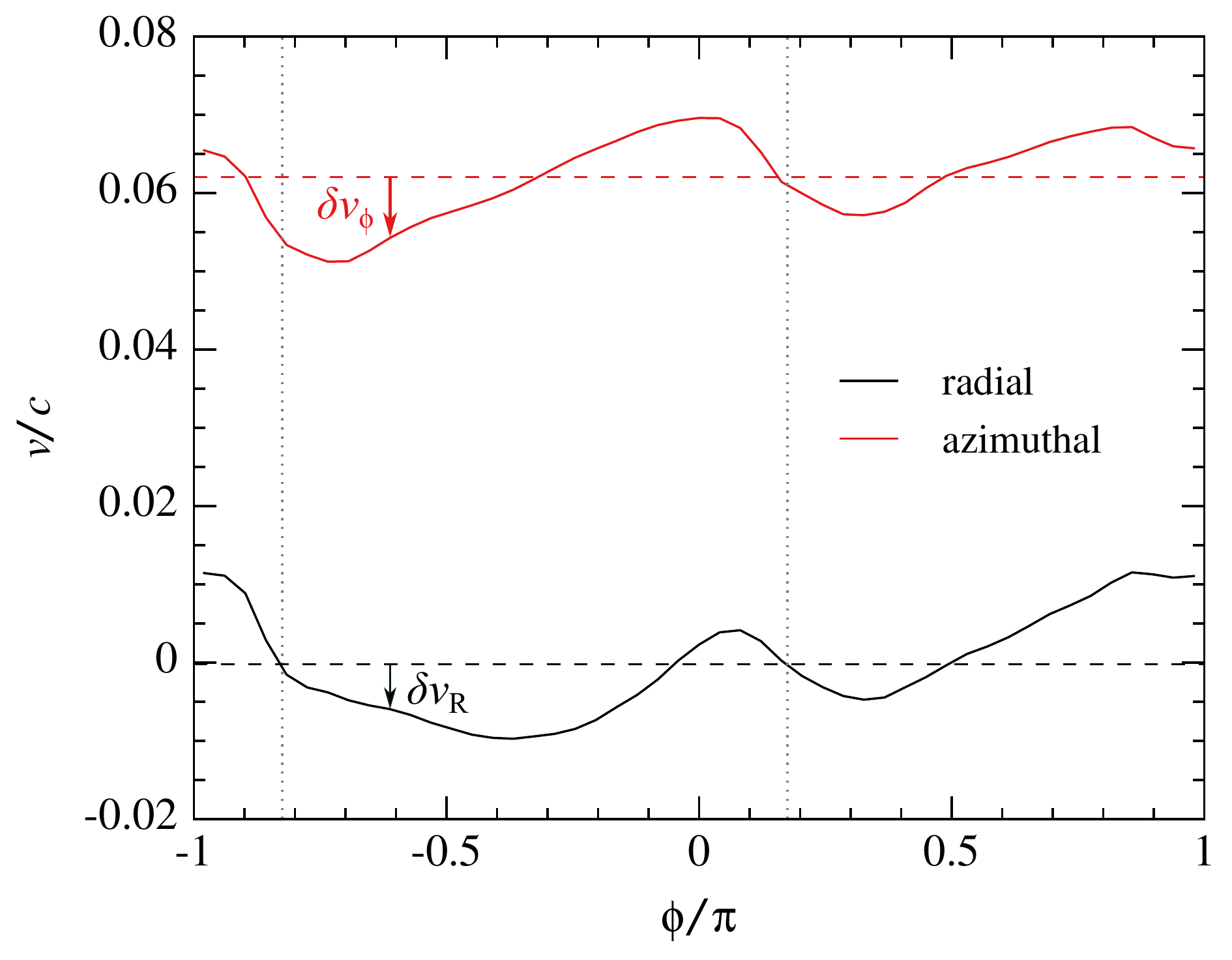}
\caption{Azimuthal profiles of the radial (solid black line) and azimuthal (solid red line) velocity components $v_{\rm R}$ and $v_{\phi}$ time-averaged in the interval $0.4 \leq t/\tmin\leq 0.5$ for the gas contained in a small annulus centred around $R = 10 \rt$ and within a polar angle $\pi/8$ from the equatorial plane. The horizontal dashed lines show the azimuthal average $\langle v_{\rm R} \rangle$ and $\langle v_{\phi} \rangle$ of these profiles. The difference between the two curves represented by the arrows correspond to the perturbed velocity components $\delta v_{\rm R} = v_{\rm R} - \langle v_{\rm R} \rangle$ and $\delta v_{\phi} = v_{\phi} - \langle v_{\phi} \rangle$. The vertical grey dotted lines represent the azimuthal angles of the two spirals obtained from the fitted red dashed curves of Fig. \ref{fig:dispersion}.}
\label{fig:vvsphi}
\end{figure}

\subsubsection{Spiral shocks}
\label{sec:spiral-shocks}

The disc displays two spirals of higher density that form around $t/\tmin = 0.3$ and remain present until $t/\tmin \approx 0.6$. They can be seen by looking at the density distribution displayed in Fig. \ref{fig:dispersion} for $t/\tmin = 0.4$ (see also the snapshots in Fig. \ref{fig:density-time-xy}). In particular, the inset shows a zoom-in on a part of the most evident left spiral with gas density rescaled to make it more visible with the density only increasing by a factor of $\sim 2$ from blue to orange. The external part of this feature coincides with the location where the filament of bound matter joins the disc outer edge, as discussed in Section \ref{sec:secondary}. Although clearly present, the spiral on the opposite side is weaker and more intermittent. This density pattern remains at the same location around the black hole with no rotation. We identify these overdensities as spiral shocks driven by the external perturbation caused by the stream of injected bound gas that continuously strikes the outer edge of the disc on the left-hand side. This is supported by the fact that the gas dissipates part of its kinetic energy when crossing the density features, proving that they are shocks.

Spiral shocks can be localized using the dispersion relation of the density wave from which they develop that is given by
\be
\left[m(\Omega - \Omega_{\rm p})\right]^2 = \kappa^2 + c^2_{\rm s} k^2,
\label{eq:dispersion-relation}
\ee
\citetext{equation (6.55) of \citet{binney2008} without considering self-gravity} where $\Omega$ and $\Omega_{\rm p}$ are the gas and pattern angular frequencies while $\kappa^2 = R^{-3} \diff (R^4 \Omega^2)/\diff R$ represents the squared epicyclic frequency. Here, $k$ and $m$ denote the radial and azimuthal wave numbers of the perturbation, respectively. We adopt $\kappa \approx \Omega$ that corresponds to a gas angular momentum profile close to Keplerian. Even though it is less accurate in the outer parts of the disc where the orbits are significantly sub-Keplerian (lower panel of Fig. \ref{fig:energy-angmom}), this assumption only weakly affects our conclusion on the location of the spiral shocks. Additionally, $\Omega_{\rm p} \approx 0$ because the perturbation does not rotate around the black hole. Equation \eqref{eq:dispersion-relation} then leads to $k \approx \Omega (m^2-1)^{1/2}/c_{\rm s}$ such that the location of the density wave is given by $\diff \phi/\diff R = -k/m \approx -\mathcal{M} (m^2-1)^{1/2} / (R m)$ where $\mathcal{M} = R \Omega/c_{\rm s} $ denotes the Mach number and $\phi$ is the azimuthal angle measured from the positive $\vect{e_{\rm x}}$ direction and increasing in the counter-clockwise way for the point of view adopted in Fig. \ref{fig:dispersion}. The solution describes a trailing spiral of equation 
\be
R \approx R_0 \, e^{-A(\phi - \phi_0)},
\label{eq:spiral}
\ee
where $A =  m / (\mathcal{M} (m^2-1)^{1/2})$ while $R_0$ and $\phi_0$ are the inner radius and azimuthal angle of the feature. For a given value of $m$, the shape of the resulting spiral is determined by the Mach number. Increasing it decreases $A$ that makes the spiral more tightly wound.

Following \citet{ju2016}, we fit the overdensities visible in the disc using equation \eqref{eq:spiral} assuming $m=2$ since two spirals are visible. The result is shown in Fig. \ref{fig:dispersion} where the two dashed red lines correspond to the fitted spirals that use $R_0 = \rt$ and $\phi_0 = 0$ while the Mach number is set to $\mathcal{M} = 1.3$ that is similar to that measured from the simulation, which is expected since $\mathcal{M} \approx H/R \approx 1$ in accretion discs \citep{frank2002}. The quality of this fit is satisfying considering the violent gas motion and large variations of its properties within the forming disc. It therefore favours our interpretation that the density pattern seen in the simulation is due to spiral shocks.

\subsubsection{Viscosity}
\label{sec:viscosity}

The gas passing through the spiral shocks experiences a change in velocity that can be seen directly from the inset of Fig. \ref{fig:dispersion} whose black arrows display both the magnitude and direction of the velocity field. To evaluate this effect more quantitatively, we compute the perturbations in the radial and azimuthal components of the velocity compared to the background flow from $\delta v_{\rm R} = v_{\rm R} - \langle v_{\rm R} \rangle$ and $\delta v_{\rm \phi} =  v_{\rm \phi} - \langle v_{\rm \phi} \rangle$, where the brackets represent time and azimuthal averages. Fig. \ref{fig:vvsphi} shows the evolution of the radial (black solid line) and azimuthal (red solid line) velocity components time-averaged in the interval $0.4 \leq t/\tmin \leq 0.5$ as a function of azimuthal angle for the gas contained in a small annulus centred at radius $R = 10\rt$ and within a polar angle $\pi/8$ from the equatorial plane. The horizontal dashed lines of the same color denote the azimuthal averages of these quantities, namely $\langle v_{\rm R} \rangle$ and $\langle v_{\phi} \rangle$. The velocity perturbations $\delta v_{\rm R}$ and $\delta v_{\phi}$ are simply the difference between these two curves as indicated by the black and red arrows. The vertical grey dotted lines represent the azimuthal angles of the two spirals obtained from the fitted red dashed curves of Fig. \ref{fig:dispersion}. Both radial and azimuthal velocities decrease around these locations with the associated velocity perturbations taking negative values, that is $\delta v_{\rm R} <0$ and $\delta v_{\rm \phi} <0$. The most prominent spiral feature on the left causes the largest and most extended velocity variations while the weaker overdensity on the opposite side leads to smaller and more localized perturbations.

It has long been known that spiral shocks can lead to outward angular momentum transport in astrophysical discs. The reason is that the components of the induced velocity perturbations are correlated such that $\langle \delta v_{\rm R} \delta v_{\rm \phi} \rangle >0$, as can be clearly seen from Fig. \ref{fig:vvsphi}. To evaluate the amount of associated inward motion, we compute the viscous parameter $\alpha = T_{\rm R \phi}/\langle P \rangle$, where $\langle P \rangle$ denotes the average pressure while $T_{\rm R \phi} = \langle \rho \delta v_{\rm R} \delta v_{\rm \phi} \rangle$ is the Reynolds stress tensor. We find that $\alpha \approx 0.02$ within $R \lesssim 15 \rt$ that corresponds to the location of the disc, thus confirming that the velocity perturbations are correlated at all distances in this region. The corresponding stress results in an inflow of gas inside the disc at a rate
\be
\dot{M}^{\rm st}_{\rm in}  \approx \alpha \langle \rho c_{\rm s} \rangle R^2 \omega,
\label{eq:mdotst}
\ee
evaluating the mass flux from $\langle \rho v_{\rm r} \rangle \approx \alpha \langle \rho c_{\rm s} \rangle$ where the radial velocity is obtained from $v_{\rm r} \approx \nu/R  \approx \alpha c_{\rm s}$ that evaluates the kinematic viscosity from the prescription $\nu \approx \alpha c_{\rm s} H$ \citep{frank2002} and uses the fact that $H/R \approx 1$ in the disc. Here, $\omega$ denotes the solid angle of the disc given by $\omega = 4 \pi \sin(\pi/8)$ since this region contains by definition the gas located within a polar angle $\pi/8$ from the mid-plane. Using the radial dependence of the parameter $\alpha$ derived above, we calculate the inflow rate induced by the stress as a function of radius. The resulting profile time-averaged in the interval $0.4 \leq t/ \tmin \leq 0.5$ is shown with a blue dash-dotted line in Fig. \ref{fig:mdotinvsr} to compare it with that obtained directly from the gas motion inside the disc (blue dashed line). At $R \lesssim 10 \rt$, the two rates roughly agree with $\dot{M}^{\rm st}_{\rm in} \approx \dot{M}_{\rm in}$, demonstrating that gas inflow is induced by the passage of the debris through the spiral shocks where their angular momentum is transported outward. As anticipated in Section \ref{sec:inflow}, the inward motion at larger radii is instead dominated by matter falling from large distances and penetrating through the disc with a residual radial velocity.

Because magnetic fields are not included in our simulation, we cannot capture the development of the magneto-rotational instability \citetext{MRI, \citealt{balbus1991}}. The gas orbital period in the disc is of order $t/\tmin \approx 0.01$ within $R\leq 5\rt$, which implies that MRI rapidly reaches saturation there. The resulting angular momentum transport is likely to significantly enhance the viscosity compared to that computed above and only due to spiral shocks. This point is discussed further in Section \ref{sec:fields}. A definitive evaluation of the amount of viscosity inside the disc requires magneto-hydrodynamics simulations of the process, which we defer to future investigations.

\begin{figure}
\centering
\includegraphics[width=\columnwidth]{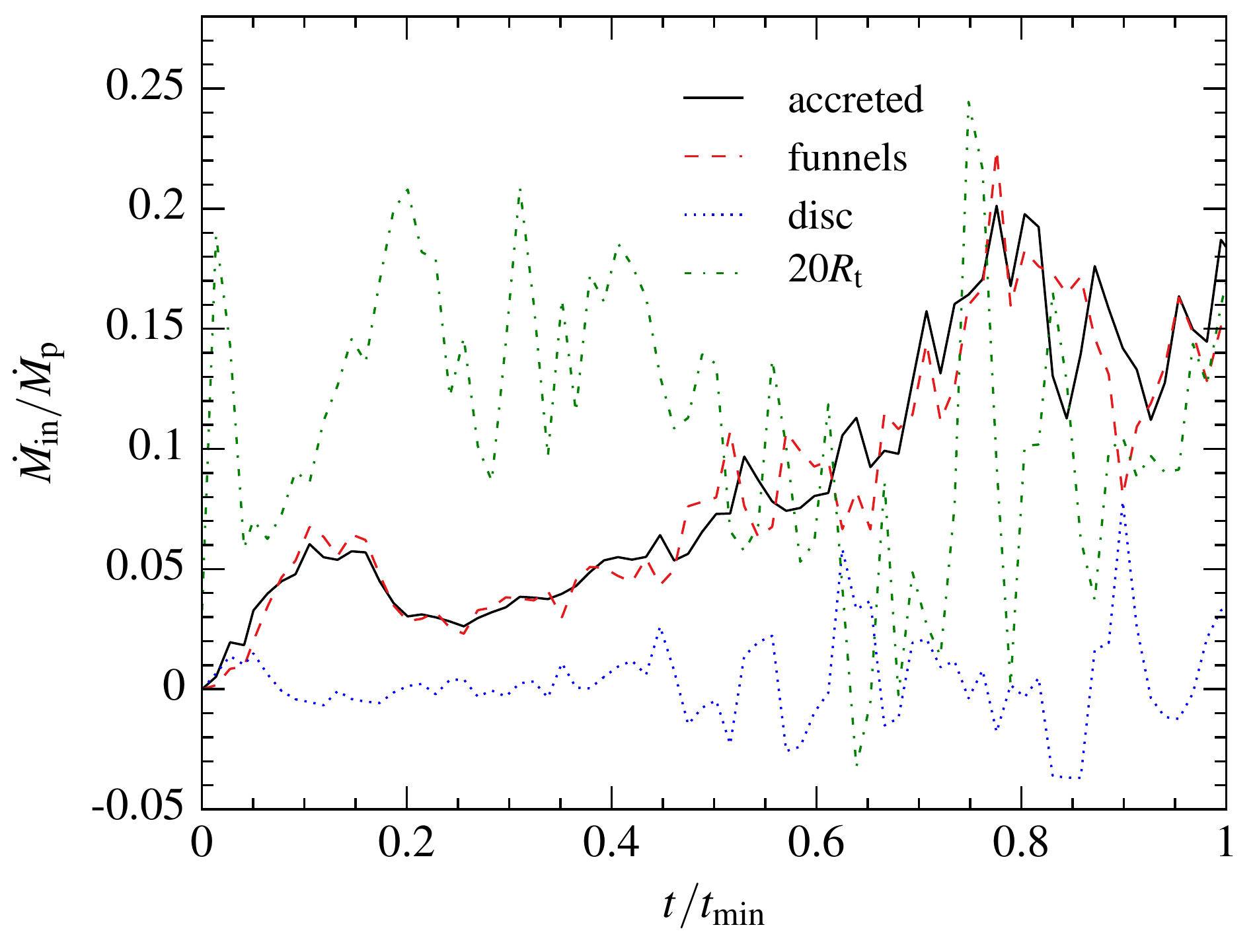}
\caption{Evolution of the accretion rate onto the black hole computed from the gas entering the accretion radius at $R_{\rm acc} = 6 \rg$ (solid black line). The inflow rates through the funnels (red dashed line) and the disc (blue dotted line) at radius $R=3 \rt \approx 45 \rg$ are also shown.  The disc is defined as the region within a polar angle $\pi/8$ from the equatorial plane while the funnels occupy the rest of the solid angle. The green dot-dashed line represents the total inflow rate measured at $R=20 \rt$ that corresponds to matter joining the disc-funnel region from any direction.}
\label{fig:mdotinvst}
\end{figure}

\subsubsection{Accretion}
\label{sec:accretion}

A fraction of the inflowing gas can make it close enough to be accreted onto the black hole. In our simulation, a fluid element is accreted if it crosses the accretion radius located at $\racc = 6\rg$. The evolution of the corresponding accretion rate is shown with a black solid line in Fig. \ref{fig:mdotinvst} that also displays the inflow rates measured at a distance of $R = 3\rt \approx 45 \rg$ through the funnels (red dashed line) and the disc (blue dotted line). Accretion onto the black hole is dominated by gas moving along the funnels while inward motion inside the disc provides only a small contribution, as anticipated in Section \ref{sec:inflow}. At early times, the mass accretion rate exhibits a small peak at $\mdotacc \approx 0.06 \, \mdotp = 1.5 \mdotedd$ due to the infall of debris that has recently been injected and can rapidly reach the inner regions. This is possible since the disc acting as a centrifugal barrier has not yet fully formed, as can be seen in Fig. \ref{fig:density-time-yz} at $t/\tmin=0.1$. Matter accumulates in the inner region as evidenced by the green dash-dotted line showing the total inflow rate at $R = 20 \rt$, which is larger than that measured at lower radii. This accumulation leads to an increase of the accretion rate to $\mdotacc \approx 0.2 \,\mdotp = 5 \mdotedd$ that largely compensates the outer inflow of matter from $t/\tmin \approx 0.5$. As a result the mass of non-accreted gas enclosed inside $R = 20 \rt$ settles to $M_{\rm enc} \approx 0.02 \mstar$ that remains roughly constant until the end of the simulation at $t/\tmin = 1$. 

One might expect that most of the gas accreted in ballistic infall through the funnels retains a significant fraction of its orbital energy when disappearing inside the black hole. For example, the gas parcel following the blue trajectory in Fig. \ref{fig:trajectories-zy} appears to get accreted in a single ballistic infall starting around $R_{\rm ff} \approx  10 \rt = 150 \rg$ conserving its orbital energy $\epsilon_{\rm ff}$ during the process. If this were true, this fluid element would dissipate only a small fraction $\epsilon_{\rm ff}/\epsilon_{\rm c} \approx R_{\rm acc} / R_{\rm ff} = 0.04$ of the maximum possible energy $\epsilon_{\rm c}$ of a circular orbit at $R_{\rm acc}$. However, we do not find that this is the case in our simulation. Instead, the large majority of the mass is accreted with an orbital energy similar to $\epsilon_{\rm c}$ after it has dissipated its original excess energy by shocks taking place in the inner region of the disc. This behaviour is different from that found by \citet{sadowski2016} who put forward this low dissipation efficiency as a possible origin for the low radiative efficiency of the disc. Instead, our simulation shows that accretion is often associated to an almost optimal thermal energy injection into the surrounding gas. Note that it is nevertheless possible that the heated gas itself enters the accretion radius before it is able to significantly cool, thus again reducing the radiative efficiency of accretion, as further examined in Section \ref{sec:radiation}.

\begin{figure}
\centering
\includegraphics[width=\columnwidth]{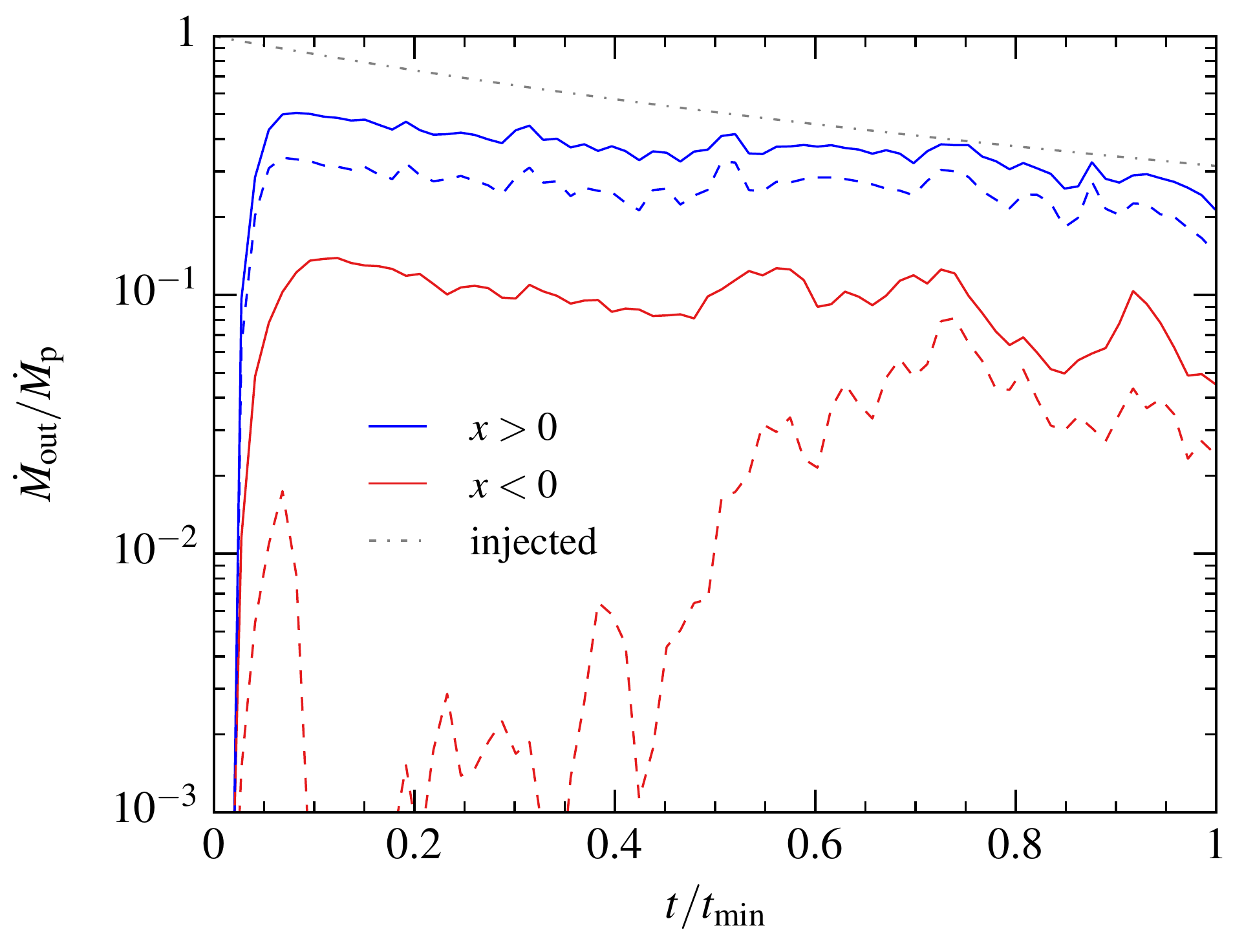}
\caption{Evolution of the mass outflow rate through a sphere located at $R = 50 \rt$ from the black hole considering the gas crossing the right hemisphere at $x>0$ (blue solid line) and the left one at $x<0$ (red solid line). The dashed lines of the same colors show these two components of the outflowing rate selecting only the gas with positive Bernoulli number $\mathcal{B}>0$. The grey dot-dashed line represents the injection rate computed analytically from equation \eqref{eq:injection-rate}.}
\label{fig:mdotoutvst}
\end{figure}

\begin{figure}
\centering
\includegraphics[width=0.96\columnwidth]{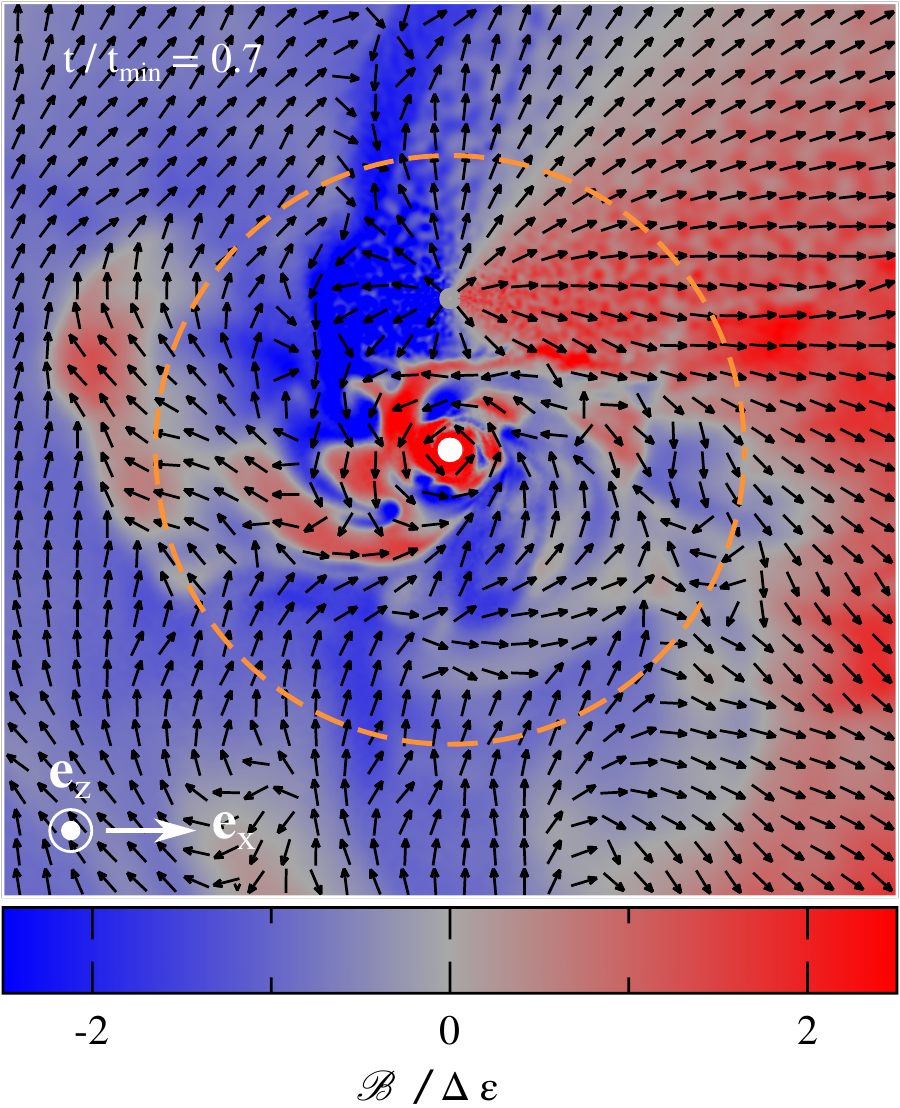}
\caption{Snapshot showing the Bernoulli number $\mathcal{B} = \epsilon +4P/\rho$ of the gas at $t/\tmin = 0.7$. The regions with $\mathcal{B} >0$ are shown in red while those displayed in blue have $\mathcal{B}<0$. The black arrows of the same size represent the direction of the velocity field while the orange dashed circle is located at $R=50\rt$ where the outflow rates of Fig. \ref{fig:mdotoutvst} are computed. The grey circle indicates the location of the injection point while the white dot represents the black hole.}
\label{fig:bernoulli}
\end{figure}

\subsection{Outflow}
\label{sec:outflow}

As explained in Section \ref{sec:simulations}, the self-crossing shock drives a massive outflow. Out of this injected mass, about $33 \%$ is initially unbound from the black hole and gets ejected near the $\vect{e_{\rm x}}$ direction. An additional source of outflow relates to secondary shocks that can launch part of the initially bound debris to large distances with a fraction being unbound. In fact, unbinding most of the matter is energetically possible if a small fraction of mass circularizes into a Keplerian orbit near $\rt$ and transfers its dissipated orbital energy to the rest of the gas. We may therefore expect that this supplementary heating significantly contributes to the outflowing gas motion.

We compute the outflow rate through a sphere located at a distance of $R = 50 \rt$ from the black hole from $\mdotout = \Delta M_{\rm out} / \Delta t$ where $\Delta M_{\rm out}$ is the mass crossing this radius going outward within a time $\Delta t$. Fig. \ref{fig:mdotoutvst} displays its evolution for the gas reaching this distance inside the right (blue solid line) and left (red solid line) hemispheres at $x>0$ and $x<0$, respectively. Both outflow rates decline following the injection rate shown with the grey dot-dashed line and analytically computed from equation \eqref{eq:injection-rate}. This is because most of the injected mass reaches distances of order $\amin \approx 85 \rt > 50 \rt$ as it goes away from the self-crossing point in a quasi-spherical outflow. The contribution from the right hemisphere is larger by a factor of a few than that from the left one because the gas has a net velocity along the $\ex$ direction.

\begin{figure}
\centering

\includegraphics[width=\columnwidth]{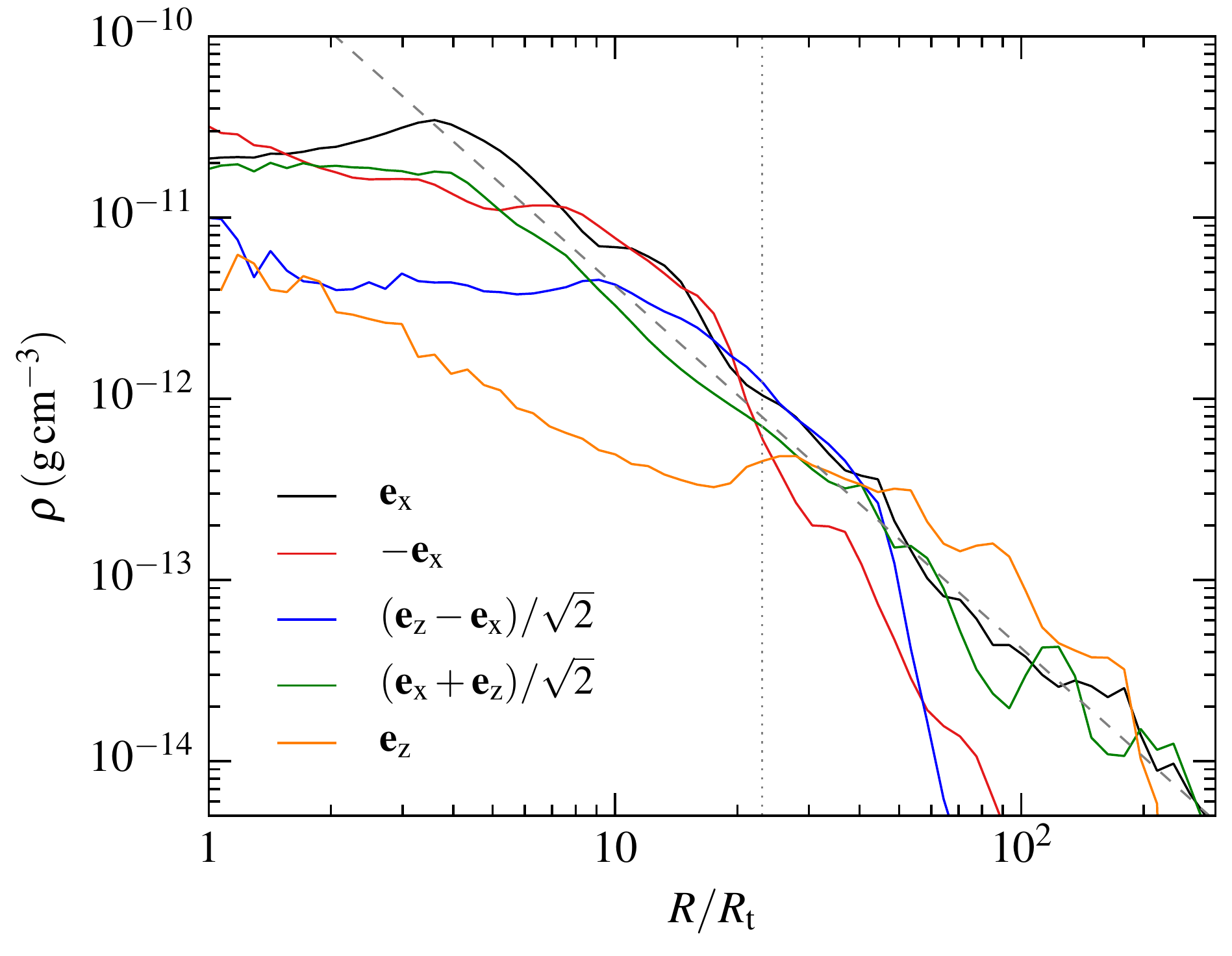}
\caption{Density profiles at $t/\tmin = 0.5$ along the $\ex$ (black line), $-\ex$ (red line), $(\ez-\ex)/\sqrt{2}$ (blue line), $(\ex+\ez)/\sqrt{2}$ (green line) and $\ez$ (orange line) directions. The dashed grey line displays the analytical description of equation \eqref{eq:density} while the dotted grey curve indicates the position of the trapping radius computed from equation \eqref{eq:trapping} using $\mdotout = 0.5 \mdotp$, $v_{\rm out} = 0.01 c$ and $v = 0.03 c$.}
\label{fig:rhovsr}
\end{figure}

Only a fraction of this outward-moving gas is unbound from the black hole. This can be seen from Fig. \ref{fig:mdotoutvst} that also shows the outflowing rates through the right (blue dashed line) and left (red dashed line) hemispheres considering only the gas with positive Bernoulli number $\mathcal{B}>0$. Even though we later refer to this gas as being unbound, we highlight that it is not necessarily the case if this matter experiences interactions at larger radii or radiates a significant fraction of its internal energy (see Section \ref{sec:radiation}). Both of these processes are likely to result in $\mathcal{B} < 0$ after the gas has crossed the sphere such that it in fact remains bound to the black hole. The component crossing the right hemisphere is dominated by gas originally unbound by the self-crossing shock that is confined to the region with $x>0$. It remains close to the initial value of $33 \%$ of the injection rate at all times. The unbound outflow rate through the left hemisphere cannot be due to gas coming directly from the self-crossing shock since this matter does not reach $x<0$. It therefore has necessarily gained its energy from interactions happening near the black hole. The spike at $t/\tmin \lesssim 0.1$ is produced by gas involved in the first secondary shocks that gains sufficient energy to escape near the $-\ey$ direction as can be seen from Fig. \ref{fig:density-time-yz} at $t/\tmin = 0.03$. Later on, the large majority of the outflow through this hemisphere is bound due to the confinement by later-arriving debris. This unbound outflow rate increases again from $t/\tmin \approx 0.5$ due to the enhanced dissipation occurring inside the disc after it has settled. Nevertheless, this unbound component remains at all times about an order of magnitude lower than that coming directly from the self-crossing point.

The energy content of the outward-moving matter can be further analysed by looking at Fig. \ref{fig:bernoulli} that shows the distribution of Bernoulli number for the gas contained in a slice parallel to the equatorial plane at $t/\tmin = 0.7$. The black arrows of the same size represent the direction of the velocity field while the orange dashed circle is located at $R=50 \rt$ where the outflow rates of Fig. \ref{fig:mdotoutvst} are evaluated. Most of the gas with $\mathcal{B} >0$ (red) is produced by the self-crossing shock and able to cross this sphere at $x>0$. Unbound gas is also present near the black hole due to heating by secondary shocks as expected from the upper panel of Fig. \ref{fig:energy-angmom}. Some of it forms pockets that are able to move outward to larger radii with $x<0$. Nevertheless, most of the matter in this region remains bound with $\mathcal{B} <0$ (blue) after being injected.

The outflowing matter reaches large distances from the black hole that produces a large-scale envelope around it. Fig. \ref{fig:rhovsr} shows density profiles at $t/\tmin = 0.5$ for different radial directions obtained by considering the gas inside cones of opening angle $\pi/8$ and centred along the unit vectors $\ex$ (black line), $-\ex$ (red line), $(\ez-\ex)/\sqrt{2}$ (blue line), $(\ex+\ez)/\sqrt{2}$ (green line) and $\ez$ (orange line). For $R \lesssim 20 \rt$, the density clearly decreases for lower polar angles, being the highest along the $\pm\ex$ ($\theta = \pi/2$) directions and the lowest along the $\ez$ ($\theta = 0$) direction. This is due to the formation of the funnels that evacuates the polar regions during disc formation. Around $R \approx 30 \rt$, the density is similar along all directions that reflects the quasi-spherical nature of the outflow. In this region, the densities roughly agree with the grey dashed line that displays the analytical profile corresponding to a steady-state spherical wind given by
\be
\begin{split}
\rho & = \frac{\dot{M}_{\rm out}}{4 \pi R^2 v_{\rm out}} \\ & \approx 2 \times 10^{-13} \gcm3 \left( \frac{R}{50 \, \rt} \right)^{-2} \left( \frac{v_{\rm out}}{0.01 c} \right)^{-1} \left( \frac{\dot{M}_{\rm out}}{0.5 \, \mdotp} \right),
\end{split}
\label{eq:density}
\ee
where the outflow rate is set to $\mdotout = 0.5 \mdotp$ and the gas outward velocity to $v_{\rm out} = 0.01 c$ as estimated from the simulation. The density profiles keep following this analytical prescription at larger radii for the $\ex$, $(\ex+\ez)/\sqrt{2}$ and $\ez$ directions corresponding to $x\geq 0$ since the gas in this region is either weakly bound or unbound and can therefore reach large distances. However, the density declines faster for $R \gtrsim 40 \rt$ for the $-\ex$ and $(\ez-\ex)/\sqrt{2}$ directions that correspond to $x<0$. This is because the gas in this region is the most bound and has therefore the lowest apocenter distances that prevent the gas from moving to large radii. Overall, the properties of this outflow are consistent with the description adopted by \cite{metzger2016} who assume that most of the matter gets launched in a wind after falling back near the black hole. However, we emphasize that the outflow is not steady-state but possesses instead complex energy and density distributions as can be seen from Fig. \ref{fig:bernoulli}.

\subsection{Heating and radiation} 
\label{sec:radiation}

A large amount of heat is injected by secondary shocks. Some of this thermal energy excess can be radiated since photons are transported outward by a combination of advection and diffusion until some of them eventually leak out entirely from the system. While it incorporates heating by shocks, our simulation does not capture photon diffusion and assumes instead that radiation is advected with the flow at every location by adopting an adiabatic equation of state with pressure dominated by radiation. Nevertheless, we perform simple post-processing calculations both to evaluate the validity of this assumption and approximately estimate the radiative luminosity that emerges during the disc formation process from the shock-heating obtained from the simulated gas evolution.

\begin{figure}
\centering
\includegraphics[width=\columnwidth]{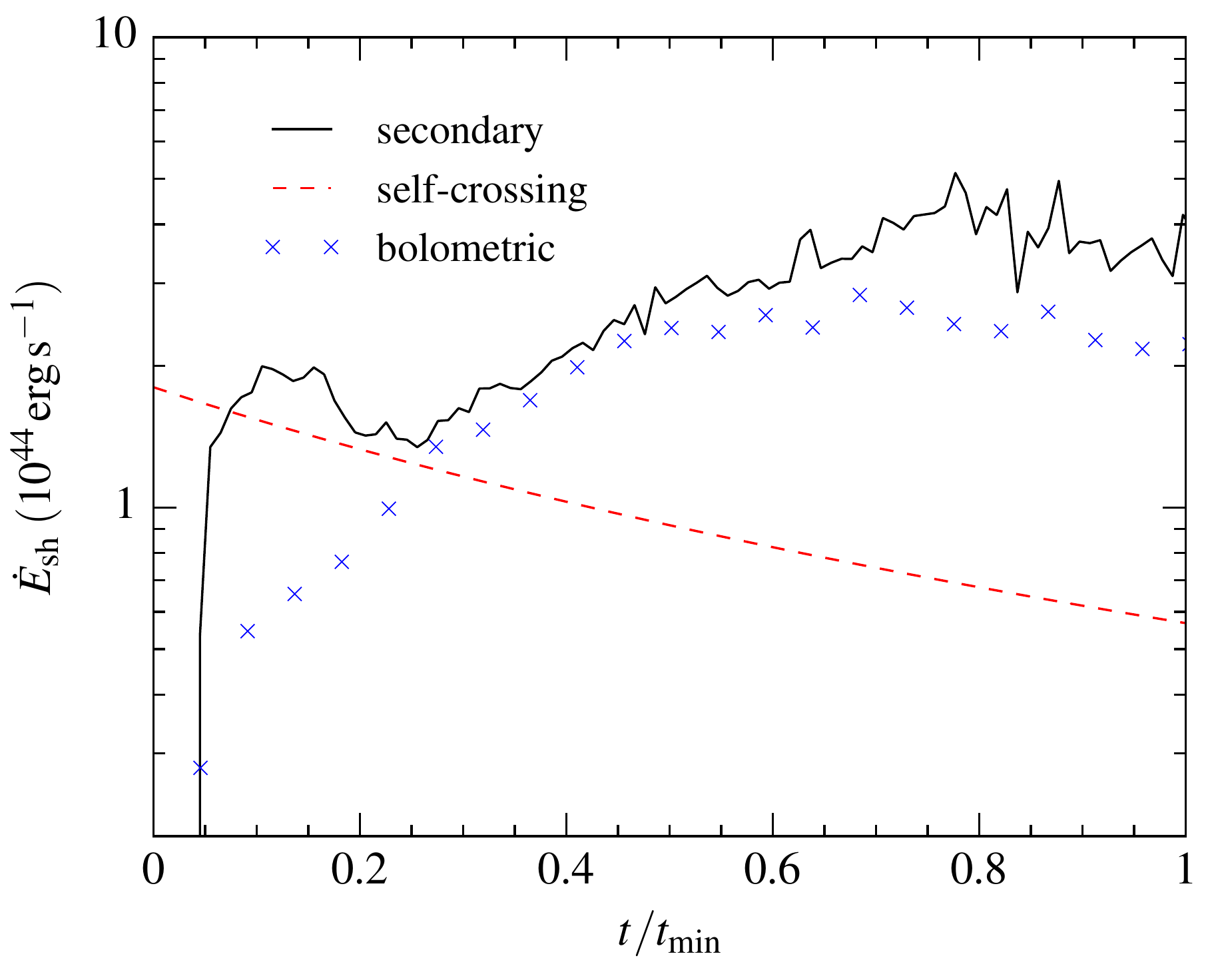}
\caption{Evolution of heating rate produced by the secondary shocks occurring in our simulation (black solid line) compared with that of the initial self-crossing shock (red dashed line) computed analytically from $\dot{M}_{\rm sc} v^2_{\rm e}/2 \approx 2 \times 10^{44} \ergpers (1+t/\tmin)^{-5/3}$. The blue crosses show an estimate of the bolometric luminosity obtained post-processing by integrating the diffusion flux on the trapping surface.}
\label{fig:luminosity}
\end{figure}

The heating rate $\dot{E}_{\rm sh}$ resulting from all the shocks experienced by the gas can be computed directly from our simulation using equation (42) of \citet{price2018} thanks to the shock detector implemented in our code. Its evolution is shown with a black solid line in Fig. \ref{fig:luminosity} along with that of the initial self-crossing shock (red dashed line) computed analytically from $\dot{M}_{\rm sc} v^2_{\rm e}/2 \approx 2 \times 10^{44} \ergpers (1+t/\tmin)^{-5/3}$. The heating rate due to secondary shocks exhibits a small bump to $\dot{E}_{\rm sh} \approx 2 \times 10^{44} \ergpers$ visible at $t/\tmin \approx 0.1$ related to the early inflow of gas near the black hole also visible in the net inflow rate along the funnels at $R = 3\rt$ shown with a dashed red line in Fig. \ref{fig:mdotinvst}. As the disc builds up, more mass gets funnelled in the inner regions and shock-heated that leads to an increase of the associated heating rate to its maximal value of $\dot{E}_{\rm sh} \approx 4 \times 10^{44} \ergpers \approx  L_{\rm Edd}$ at $t/ \tmin \approx 0.7$. From that moment, it remains roughly constant until the end of the simulation as the disc density settles while the rate of injected mass drops. The heating rate due to the initial self-crossing shock drops following the decline in the fallback rate to become about an order of magnitude lower than that produced by secondary shocks. This is because most shocks in our simulation take place closer to the black hole than the intersection point where relative velocities are larger leading to more heating.

We now want to estimate the fraction of this generated heat that participates to the emerging luminosity. Close to the black hole where densities are the largest, numerous scatterings by free electrons prevent photons from moving with respect to the gas. At larger radii, the decreased density allows radiation to decouple from the matter and diffuse away. The transition occurs at the trapping radius where the diffusion timescale $t_{\rm diff} \approx \tau R/c$ equals the dynamical timescale $t_{\rm dyn} \approx R/v$. Here, $\tau$ denotes the optical depth while $v$ is the gas velocity. This equality yields $\tau \approx c/v$ such that the trapping radius $R_{\rm tr}$ is given by
\be
\int_{R_{\rm tr}}^{+\infty} \rho \kappa_{\rm s} \diff R = c/v(R_{\rm tr}),
\label{eq:trapping}
\ee
where $ \kappa_{\rm s} = 0.34 \cm2g$ denotes the electron scattering opacity and the velocity on the right-hand side is evaluated at the trapping radius. We compute this critical distance by performing the integral of equation \eqref{eq:trapping} from outside in until it reaches the local ratio of speed of light to gas velocity. This calculation is carried out considering different radial paths corresponding to given values of $\theta$ and $\phi$ that allows us to specify the trapping surface, from which photons are able to diffuse from the gas.

\begin{figure}
\centering
\includegraphics[width=\columnwidth]{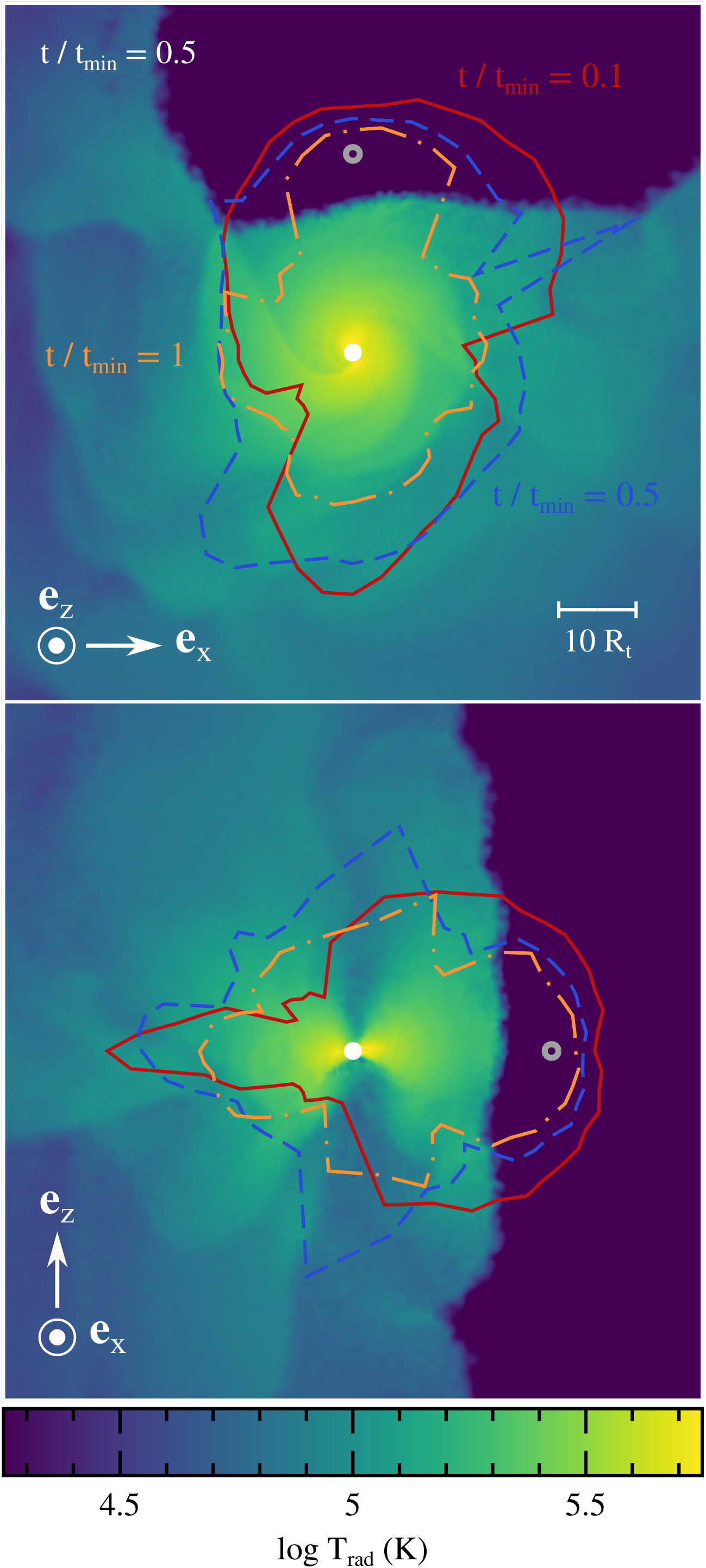}
\caption{Snapshots showing the gas distribution at $t/\tmin= 0.5$ along slices parallel (upper panel) and perpendicular (lower panel) to the disc mid-plane. The colours correspond to the radiation temperature whose value is indicated in the colour bar, increasing from purple to yellow. Overplotted lines represent the location of the trapping surface defined in equation \eqref{eq:trapping} at $t/\tmin=$ 0.1 (red solid line), 0.5 (blue dashed line) and 1 (orange dash-dotted line). The grey circle indicates the location of the injection point while the white dot represents the black hole.}
\label{fig:temperature}
\end{figure}

The trapping surface is shown in Fig. \ref{fig:temperature} at different times with lines overplotted on the gas distribution of radiation temperature evaluated at $t/\tmin = 0.5$ along  slices parallel (upper panel) and orthogonal (lower panel) to the equatorial plane and computed from $T_{\rm rad} =  (e/a)^{1/4}$ where $e$ is the internal energy density while $a = 7.6 \times 10^{-15} \, \rm erg \, cm^{-3} \, K^{-4}$ denotes the radiation constant. The red solid line corresponds to $t/\tmin = 0.1$ when the trapping surface is growing primarily along the $-\ey$ direction as freshly-injected gas reaches this location, as can be seen directly from the corresponding density snapshots of Figs. \ref{fig:density-time-xy} and \ref{fig:density-time-yz}. The maximal extent of this region is reached around $t/\tmin = 0.5$ (blue dashed line), from which it starts receding inward as the density drops due to the decline in the injection rate. As a result, the trapping surface is the smallest at $t/\tmin = 1$ (orange dot-dashed line). The trapping radius is only weakly dependent on the radial path considered because the gas envelope surrounding the black hole has an approximately spherical density structure as explained in Section \ref{sec:outflow}. In particular, we note that the trapping surface is not affected by the presence of the funnels at low radii. The average value of the trapping radius can be recovered by plugging the analytical density profile of equation \eqref{eq:density} into equation \eqref{eq:trapping} that yields
\be
R_{\rm tr} \approx \frac{\mdotout \kappa_{\rm s}}{4 \pi c} \frac{v}{v_{\rm out}} = 23 \rt \left( \frac{\mdotout}{0.5 \mdotp} \right) \left( \frac{v}{0.03 c} \right) \left( \frac{v_{\rm out}}{0.01 c}\right)^{-1},
\ee
where the total velocity of $v = 0.03 c$ is evaluated based on the simulation near the trapping radius at $t/\tmin = 0.5$. This value is indicated by a vertical grey dotted line on the density profiles of Fig. \ref{fig:rhovsr}. At all times, the trapping radius encloses the region where most of the interactions between debris take place, which implies that our assumption of adiabaticity is valid for most of the hydrodynamics described in the above sections. One exception concerns the unbound outflow that we define in Section \ref{sec:outflow} based on the conservation of the Bernoulli number that assumes adiabaticity. Taking into account energy loss through radiation outside the trapping radius would reduce the unbound fraction compared to what we estimated. Similarly, the radiation temperature is computed accurately from the inner regions where $T_{\rm rad} \approx 10^6 \rm K$  to the trapping surface where it has decreased to $T_{\rm rad} \approx 10^5 \rm K$ due to the adiabatic losses in the expansion driven by radiation pressure. However, our assumption of advective transport breaks down outside the trapping radius where photons decouple from the gas. This implies that our calculation of the radiation temperature is not self-consistent in these outer parts as discussed further in Section \ref{sec:emerging}.

The trapping radius is only marginally larger than the outer edge of the disc where most of the dissipation takes place. We therefore do not expect significant adiabatic losses as radiation is advected to this surface before it diffuses outward. To evaluate this effect, we show in Fig. \ref{fig:luminosity} with blue crosses the evolution of the emerging bolometric luminosity obtained by integrating on the trapping surface the diffusion flux given by $F_{\rm dif} =  c |\nabla e| / (3\rho \kappa_{\rm s}) \approx  \xi e c/\tau$ \citetext{equation (I.24) of \citet{shapiro1983}} computing the energy density directly from our simulation.\footnote{Radiation can be absorbed outside the trapping radius before it reaches the observer. However, these photons are rapidly re-emitted without energy loss and the bolometric luminosity is therefore unaffected by this process.} In the second term, we fix $\xi = 8/9$ as obtained from the approximate scaling $e \propto \rho^{4/3} \propto R^{-8/3}$. The bolometric luminosity being close to the heating rate from secondary shocks (solid black line) at all times confirms our expectation of small energy loss due to adiabatic expansion. Additionally, it suggests that advection and subsequent accretion of hot gas onto the black hole does not lead to a significant reduction of the energy output.

The calculations presented above to estimate how photon diffusion occurs in the forming disc are approximate and we do not pretend to have solved radiation transport in TDEs. To achieve this goal, the disc formation process has to be studied by radiation-hydrodynamics simulations, which we defer to the future.

\section{Discussion}
\label{sec:discussion}

\subsection{Resolution study}
\label{resolution}

To evaluate the influence of numerical resolution on our results, we repeat the same simulation with both a larger and lower number of particles. This is done by modifying their mass to $\mpart = 3 \times 10^{-9} \msun$ and $\mpart = 3 \times 10^{-8} \msun$ that corresponds to a higher and lower resolution compared to that $\mpart = 10^{-8} \msun$ adopted in the rest of the paper. We find that the dynamics of disc formation is quantitatively similar for the two larger resolutions with a well-defined torus forming that has the same mass $M_{\rm enc} = 0.02 \mstar$ enclosed within $R\leq 20 \rt$ at $t/\tmin = 0.5$. The gas evolution is however significantly different in the lower resolution simulation, for which the gas clusters near the black hole but does not form a coherent structure. As a result, the enclosed mass is smaller than in the two other runs with $M_{\rm enc} = 0.008 \mstar$ at the same time. This influence of resolution is similar to that obtained in the early SPH simulation by \cite{ayal2000}, who find that a disc does not form when the matter falls back near pericenter with a very low number of particles. We conclude that the resolution adopted in the simulation presented in the paper is high enough that the results are not affected by increasing it further.

\subsection{Validity of the injection procedure}

Our treatment of the self-crossing shock assumes that it is continuously joined by two streams that remain identical at all times. This approach assumes that the returning debris is unaffected by the presence of the matter that has already accumulated near the black hole. The validity of this assumption can be evaluated from the density ratio between these two components. As can be seen from Fig. \ref{fig:rhovsr}, the density profile of the gas in our simulation is accurately described by equation \eqref{eq:density}. The density of the stream can be estimated in a similar way from $\rho_{\rm s} = \mdotfb/(\pi H^2_{\rm s} v_{\rm s})$ where $H_{\rm s}$ and $v_{\rm s}$ are its width and velocity assuming cylindrical symmetry. The density ratio is then
\be
\frac{\rho_{\rm s}}{\rho} = \frac{ 4 \mdotfb v_{\rm out} R^2}{\mdotout v_{\rm s} H^2_{\rm s} } \approx 8000 \left(\frac{0.5 \mdotfb}{\mdotout}\right) \left(\frac{10 \,v_{\rm out}}{v_{\rm s}} \right) \left( \frac{0.05 \rt}{ H_{\rm s}} \right)^{2},
\label{eq:density-ratio}
\ee
evaluating it at $R \approx 5 \rt$. The outflow rate is $\mdotout \approx 0.5 \mdotfb$ at all times as evidenced by Fig. \ref{fig:mdotoutvst}. The stream moves at a speed given by $v_{\rm s} \approx (2 G \mh /5\rt)^{1/2} \approx 0.1 c$ that is about an order of magnitude larger than that of the outflowing gas, of $v_{\rm out} \approx 0.01 c$. Furthermore, the stream is able to maintain a thin profile until it reaches the vicinity of the black hole \citep{coughlin2016-structure} such that $H_{\rm s} \approx 10 \rstar \approx 0.05 \rt$. The three ratios on the right-hand side of equation \eqref{eq:density-ratio} are therefore close to unity such that the stream is significantly denser than the mass already present around the black hole when it arrives. This calculation shows that it is legitimate to neglect in first approximation the dynamical impact of the accumulated mass on the freshly-returning stream. It is nevertheless possible that this debris can be affected by hydrodynamical instabilities taking place at the boundary between the stream and the gas present in the disc or surrounding envelope. These interactions are similar to those investigated by \citet{bonnerot2016-kh} during the early evolution of the stream. When the stream becomes very tenuous at late times while mass keeps accumulating near the compact object, the ratio between the two densities is however likely to decrease, thus enhancing the impact of the surrounding matter on the trajectories of the returning debris.

\subsection{Extrapolation to other parameters}

Our choice of parameters corresponds to an intersection radius $\rint \approx 25 \rt$ that is much lower than the apocenter of the most bound debris of $2 \amin \approx 170 \rt$. This is due to the large apsidal relativistic precession experienced by the stream when reaching its pericenter at a distance of $\rp \approx 15 \rg$. As a result, the initial self-crossing shock launches the debris on a wide range of trajectories with the bound ones promptly reaching small radii. As demonstrated by \cite{lu2019}, this description likely applies to more relativistic disruptions such as those involving penetration factors $\beta > 1$ or black holes of larger masses than the one $\mh = 2.5 \times 10^6 \msun$ we consider. We therefore argue that our simulation provides at least the correct qualitative gas evolution at play during disc formation for this region of parameter space. Similar simulations have been carried out by \cite{bonnerot2016-circ} and \cite{sadowski2016} who consider the disruption of bound stars assuming a pericenter well within the tidal radius such that $\rp \approx 10 \rg$, only slightly smaller than what we use. Our results are in several respects similar to theirs, most importantly the fact that the disc forms on a timescale of a few tenths of $\tmin$.

Decreasing the black hole mass to $\mh \lesssim 10^6\msun$ while keeping $\beta =1$ reduces the apsidal precession angle such that the intersection radius rapidly increases to $\rint \gtrsim \amin$. As described by \cite{lu2019}, the initial self-crossing shock becomes weaker in this situation such that the debris that passes through it moves on a narrower range of trajectories. This type of mildly relativistic encounters has been investigated by several authors considering either initially bound stars \citep{bonnerot2016-circ,hayasaki2016-spin} or a black hole of low mass with $\mh = 500 \msun$ \citep{shiokawa2015}. These works find that about ten $\tmin$ are required for the debris to reach a significant level of circularization. It is legitimate to think that this property carries over to standard TDEs when the self-crossing shock is weak. However, extrapolating the detailed hydrodynamics found in these simulations is uncertain due to the simplifications they make. Most importantly, decreasing the stellar eccentricity results in all the debris stream being bound that artificially suppresses the tail of later-arriving matter. Considering an intermediate-mass black hole modifies the location of the self-crossing shock due to the dependence on the black hole mass of both the apsidal precession angle and the semi-major axis $\amin$. It is also likely to increase the dynamical importance of dissipation occurring during the stream compression at pericenter \citep{guillochon2014-10jh}. For these reasons, the detailed dynamics at play for this region of parameter space remains largely unclear.

\subsection{Properties of the emerging radiation}
\label{sec:emerging}

Our simulation demonstrates that the heating rate by secondary shocks quickly increases to becomes larger than that of the initial self-crossing shock by almost an order of magnitude (Fig. \ref{fig:luminosity}). Furthermore, we argue based on post-processing calculations that the bolometric luminosity is likely similar to this shock heating rate due to the proximity of the trapping radius to the newly-formed disc. This favours the interpretation of \cite{metzger2016} for the origin of the optical emission from TDEs that relies on dissipation near the tidal radius  and then radiative transport by advection and diffusion. \cite{piran2015-circ} argue instead that this emission is caused by the initial self-crossing shock when it occurs on larger scales similar to $\amin$. However, our simulation demonstrates that this initial source of heating is weaker in addition to being more susceptible to adiabatic losses \citep{lu2019}. For this reason, we argue that this option is less likely to account for observations.

We find that the radiation temperature is $T_{\rm rad} \approx 10^5 \rm K$ at the trapping surface (Fig. \ref{fig:temperature}). As photons are transported further out by diffusion, they can be absorbed and re-emitted that reduces their temperature until the thermalization radius $R_{\rm th}$ given by \citetext{equation (1.98) of \citet{rybicki1979}}
\be
\int_{R_{\rm th}}^{\infty} \rho (\kappa_{\rm a} \kappa_{\rm s})^{1/2} \diff R = 1,
\ee
where $\kappa_{\rm a}$ denotes the absorption opacity that is much lower than that $\kappa_{\rm s}$ due to scattering. In order to reproduce the optical emission detected from TDEs, this radius has to be a factor of a few larger than the trapping radius \citetext{figure 13 of \citet{lu2019}}. A more accurate determination of the spectrum could be achieved using the density distribution obtained from our simulation by carrying out post-processing radiative transfer calculations similar to those by \cite{roth2016}.

\subsection{Influence of magnetic fields}
\label{sec:fields}

Our simulation does not consider magnetic fields that the debris inherits from the star upon disruption. As mentioned in Section \ref{sec:viscosity}, including them would likely enhance angular momentum transport due to the development of the MRI compared to that produced by hydrodynamical effects alone. Furthermore, the presence of magnetic fields inside the newly-formed disc could lead to additional outflows through the \cite{blandford1982} mechanism. However, it has been found that a weak or zero net poloidal magnetic field reduces both the angular momentum transport resulting from the MRI and the strength of magnetically-driven winds \citep{hawley1995,salvesen2016,jiang2017,zhu2018}.

This situation is likely to apply to TDEs because the  magnetic field lines get aligned with the direction of stream stretching prior to disc formation that suppresses the vertical field \citep{guillochon2017-magnetic,bonnerot2017-magnetic}. For this reason, it is possible that the dynamical influence of magnetic fields is reduced compared to other accreting systems that do not lack this poloidal component. This argument is supported by the simulation of \citet{sadowski2016} who finds that the dynamics of the newly-formed disc is unaffected by the inclusion of magnetic fields even after the MRI has reached saturation. The possibility that the early accretion has an hydrodynamical origin has also been proposed by \cite{nealon2018}.

\subsection{Impact of black hole spin}

Our initial conditions assume that the self-crossing shock happens after the tip of the stream has passed pericenter and moves outward to intersect with matter still falling back. However, this initial collision can be prevented if the black hole has a non-zero spin due to Lense-Thirring precession that modifies the orbital plane of the debris during pericenter passage \citep{dai2013,guillochon2015,hayasaki2016-spin}. In this case, the stream may experience a large number of revolutions before it intersects with itself. However, \citet{guillochon2015} show that this collision is more likely between neighbouring windings that differ by only one pericenter passage. In this situation, the properties of the delayed self-crossing shock may be similar to the non-rotating case that we are simulating. One important difference relates to the presence of multiple streams that did not collide and could occupy a significant fraction of the volume directly surrounding the black hole. Interactions involving this gas could affect the dynamics of the forming disc, making it significantly more complex.

\section{Summary}
\label{sec:summary}

We carried out the first simulation of disc formation in TDEs considering realistic initial conditions valid for a star disrupted on a parabolic trajectory by a supermassive black hole. This is achieved by following the gas evolution only after it has passed through the self-crossing shock that we model by injecting SPH particles inside the computational domain according to the properties derived from the numerical study of the initial collision by \cite{lu2019}. This technique allows us to circumvent the high computational cost required to accurately capture the previous passage of the stream at pericenter. The results can be summarized as follows.

\begin{enumerate}

\item Due to the large range of trajectories followed by the gas leaving the self-crossing shock, multiple secondary shocks take place that lead to the formation of an accretion disc at $t/\tmin \approx 0.3$. This structure rotates in the direction opposite to that of the original star and retains significant eccentricities broadly distributed around $e \approx 0.2$ even after the overall gas evolution has settled. It is very thick with an aspect ratio $H/R \approx 1$ and extends to distances $R \approx 15 \rt$.

\item Pressure support is important in most of the disc such that its angular momentum profile is significantly sub-Keplerian with an almost flat radial dependence. Most of the gas in this region has a Bernoulli number $\mathcal{B}\approx 0$ that is much larger than its orbital energy due to the excess heat injected by the circularizing shocks that makes the disc prone to outflows.

\item Two overdensities develop inside the newly-formed disc that we identify as spiral shocks excited by the arrival of bound mass from the injection point that continuously strikes its outer edge. The velocity changes experienced by the debris moving through these features cause outward angular momentum transport that can account for the inward motion of gas along the mid-plane at a slow rate of $\sim 10^{-2} \mdotp = 0.25 \mdotedd$.

\item Gas inflow is dominated on small scales by matter falling along the funnels of the torus at a rate of $\mdotin \approx 0.1 \mdotp = 2.5 \mdotedd$ that can result in accretion onto the black hole. The majority of this gas stayed in the disc for several orbital periods after it has been injected from the self-crossing shock. Moreover, accretion does not happen on a single near-radial infall but after most of the gas orbital energy has been deposited in the inner region of the disc.

\item At larger distances $R \gtrsim 50 \rt$, the gas moves outward at a rate a factor of a few lower than the fallback rate. It is dominated by matter directly coming from the self-crossing shock that contains a significant fraction of unbound gas with a lower contribution from debris gaining energy from secondary shocks. The resulting density profile can be approximately described by a steady-state spherical wind coming from the black hole vicinity.

\item Secondary shocks cause gas heating at a rate peaking at $\dot{E}_{\rm sh} \approx 4\times 10^{44} \ergpers \approx L_{\rm Edd}$ that is larger than that due to the self-crossing shock by almost an order of magnitude. Post-processing calculations suggest that a large fraction contributes to the bolometric luminosity due to the proximity of the trapping radius to the outer edge of the disc that reduces adiabatic losses.
\end{enumerate}

This study and the numerical method it uses open several promising lines of research that we intend to pursue in the future. This includes the exploration of the hydrodynamics of disc formation for a wider range of parameter space, most importantly less relativistic encounters than we assume here. Additionally, we envisage simulations of this process treating in a self-consistent way radiative diffusion and angular momentum transport including magnetic fields. Major advances are foreseen that will improve our understanding of the early emission from TDEs now accessible by observational facilities.

\section*{Acknowledgments}

We thank Eliot Quataert, Sterl Phinney, Phil Hopkins, Nathan Roth, Luc Dessart, Chris White and Greg Salvesen for useful discussions. We acknowledge the use of \textsc{splash} \citep{price2007} for producing most of the figures in this paper. This research benefited from interactions at the ZTF Theory Network Meeting, partly funded by the National Science Foundation under Grant No. NSF PHY-1748958. The research of CB was funded by the Gordon and Betty Moore Foundation through Grant GBMF5076. WL was supported by the David and Ellen Lee Fellowship at Caltech.




\bibliographystyle{mnras} 
\bibliography{biblio}




\bsp	
\label{lastpage}


\end{document}